\documentclass[a4paper,11pt]{article}
\usepackage[english]{babel}
\usepackage[T1]{fontenc}
\usepackage{textcomp}
\usepackage{epsfig}
\usepackage{latexsym}
\usepackage{graphicx}
\usepackage{amsfonts,amssymb,bm}
\usepackage{color}
\usepackage{mathrsfs}

\newtheorem{proposition}{Proposition}

\newtheorem{theorem}{Theorem}
\newtheorem{corollary}{Corollary}

\newcommand{\beq}{\begin{equation}}

\newcounter{llista}

\begin{document}
\title{Matter and Ricci collineations}
\author{Josep Llosa\footnote{e-mail: pitu.llosa@ub.edu\,. Postal address: Martí i Franqués 1, E-08028 Barcelona, Spain}\\
\small Departament de Física Fonamental and
Institut de Ciències del Cosmos (ICCUB) \\
Universitat de Barcelona, Spain}

\maketitle

\begin{abstract}
The infinitesimal transformations that leave invariant a two-covariant symmetric tensor are studied. The interest of these symmetry transformations lays in the fact that this class of tensors includes the energy-momentum and Ricci tensors. 
We find that in most cases the class of infinitesimal generators of these transformations is a finite dimensional Lie algebra but also, in some cases exhibiting a higher degree of degeneracy, this class is infinite dimensional and may fail to be a Lie algebra. 

Mathematics Subject Clasification: 83C99, 53A45, 53D05, 58A10, 58A17
\end{abstract}
\section{Introduction  }
The interest in the study of symmetries in General Relativity is long-standing. Some of them, namely isometries and affine transformations and their infinitesimal counterparts, Killing vector fields and affine vector fields, are well understood since long ago \cite{Cramer}, \cite{Yano}, \cite{Hall1990}, \cite{Hall1995}. 

In the last twenty years, there has been an steady interest in curvature collineations, Ricci collineations and even matter (Einstein) collineations \cite{Bokhari1993}, \cite{Melfo1992}, \cite{Hall1991}, \cite{Carot1994}.
Their infinitesimal counterparts, namely collineation fields, are characterized by the vanishing of the Lie derivative of the curvature tensor (resp., the Ricci or the energy-momentum tensor). Collineation fields are thus an extension of the aforementioned Killing fields and affine fields in that every Killing vector field is an affine vector filed which in turn is a curvature collineation field and also a Ricci and a matter collineation field. 
However it is well known that collineation fields present new features. Indeed, contrarily to the case of Killing and affine fields, the class $\mathcal{C}$ of curvature (resp., Ricci and matter) collineation fields is a real vector space which may be infinite dimensional; this is due to the dependence on arbitrary functions, which also results in the fact that a collineation field needs not to be smooth and, as a consequence, $\mathcal{C}$ may not be a Lie algebra \cite{Hall1991}.

We shall here concentrate in infinitesimal Ricci and matter collineations. Our results are also useful in the study of curvature collineations because any of them is necessarily a Ricci collineation too.  
In our view, although most recent work on the subject the spacetime metric (from which the Ricci tensor is derived) is given a significant presence in the approach to the problem, paying attention to the metric is rather hindering than helpful.

Given a 4-manifold $\mathcal{M}$ and a smooth field of symmetric 2-covariant tensors $T$, we shall concentrate on finding the class $ \mathcal{C}_T$ of vector fields $\mathbf{X}$ such that $\mathcal{L}_{\mathbf{X}} T = 0 \,$ and try to find out whether the number of dimensions of $\mathcal{C}_T$ is finite, whether $\mathbf{X}$ is smooth and whether $\mathcal{C}_T$ is a Lie algebra.

The answer to these questions depend, but not exclusively, on the rank of $T$. Particularly, if ${\rm rank}\,T=4$, $T$ itself can be taken as a non-degenerate metric tensor and the collineation equation is actually a  Killing equation and, as it is well known \cite{Cramer}, $\mathcal{C}_T$ is a Lie subalgebra of $\mathscr{X}(\mathcal{M})$, the class of smooth vector fields,  and ${\rm dim}\,\mathcal{C}_T \leq 10$.

For ${\rm rank}\,T < 4$, we come across an assorted casuistry which depends not only on the rank of $T$ but also on the derivatives of $T$. We aim to set up a classification of the tensor fields $T$ according to its class $\mathcal{C}_T$ of collineation fields. The first variable to consider is the rank and different methods and techniques are suitable for different ranks, e. g. for rank three tensors the method is more similar to that used in studing the Killing fields whereas techniques imported from simplectic mechanics are best suited for rank one tensors. Whimsical as it could seem, the order in which the different ranks are presented here is dictated by their progressive degree of difficulty. 

The classification we obtain is rather simple if only the generic, i. e. less degenerate, cases are considered. However as the degree of degeneracy increases, an intricated mess of cases and subcases arises. This is why we clos the paper with a Summary section.  

\section{Collineation fields \label{S1}}
Let $T$ be a 2-covariant symmetric smooth tensor field on a $4$-manifold $\mathcal{M}$. A $T$-collineation field (shortly, a collineation field) is  a vector field  $\mathbf{X}$ such that
\begin{equation} \label{e1}
\mathcal{L}_{\mathbf{X}} T = 0
\end{equation}
Notice that the definiton requires that $\mathbf{X}$ is of class $\mathcal{C}^1$ at least but in general it does not guarantee that $\mathbf{X}$ is smooth.

Furthermore, as a consequence of the fact that $\left[\mathcal{L}_{\mathbf{X}},\mathcal{L}_{\mathbf{Y}} \right]  = \mathcal{L}_{[\mathbf{X},\mathbf{Y}]}$, provided that $\mathbf{X}$ and $\mathbf{Y}$ are of class $\mathcal{C}^2$ at least, it is obvious that the class of smooth collineation fields is a Lie algebra.

The case ${\rm rank}\,T = 4$ having been discussed,  and finished off, in the Introduction, we shall assume that ${\rm rank}\,T < 4$, constant. Hence  in the neighbourhood of every $p\in \mathcal{M}$ it exists a base of $\Lambda^1\mathcal{M}$, $\{\phi^a\}_{a= 1\ldots 4}$  such that\footnote{As there is no distinguished metric it is pointless to wonder whether this base is orthonormal or not.}
\begin{equation} \label{e2}
T = \eta_{\alpha\beta} \phi^\alpha \otimes \phi^\beta \,, \qquad {\rm with} \qquad \eta_{\alpha\beta}={\rm diag}(+1 \stackrel{r}{\ldots} +1,-1 \stackrel{s}{\ldots} -1)\,, \qquad r+s=m
\end{equation}
(Greek indices run from 1 to $m$, upper case indices $A, B, \ldots$ run from $m+1$ to $4$, latin indices $a, b, \ldots$ run from 1 to 4 and the summation convention is always understood unless the contrary is explicitely stated.) The 1-forms $\phi^\alpha$ being independent, we have that $\phi^1\wedge \ldots \wedge \phi^m \neq 0$ and the set of 1-forms $\{\phi^\alpha\}_{\alpha = 1 \dots m}$ is called a {\em $T$-frame}. 

Let us now expand the Lie derivatives of any $\phi^\nu$ as
% \begin{equation} \label{e3}
$\,\mathcal{L}_{\mathbf{X}}\phi^\nu = M^\nu_{\;\beta}\phi^\beta+ M^\nu_{\;B}\phi^B \,$.
%\end{equation}
Including this and (\ref{e2}) it easily follows that equation (\ref{e1}) is equivalent to
\begin{equation} \label{e4}
\mathcal{L}_{\mathbf{X}} \phi^\nu = M^\nu_{\;\beta}\phi^\beta \qquad
{\rm with} \qquad   \eta_{\alpha\nu} M^\nu_{\;\beta} = \eta_{\beta\nu}M^\nu_{\;\alpha}
\end{equation}
that is, the matrix $M^\nu_{\;\beta}$ is an $so(r,s)$-valued function on $\mathcal{M}$.

Any two $T$-frames, $\{\phi^\alpha\}_{\alpha = 1 \dots m}$ and $\{\tilde\phi^\alpha\}_{\alpha = 1 \dots m}$, are connected through an $\eta$-orthogonal transformation: 
\begin{equation} \label{MC2.1}
\tilde\phi^\alpha := R^\alpha_{\;\beta} \phi^\beta\,,  \qquad {\rm where} \qquad 
R^\mu_{\;\alpha} R^\nu_{\;\beta} \eta_{\mu\nu} = \eta_{\alpha\beta}
\end{equation}
i.e. $R^\alpha_{\;\beta}$ is a field of $\eta$-orthogonal matrices. For the sake of brevity, we shall refer hereafter to these transformations as {\em $T$-rotations}.   

The differential system associated to $T$ is 
$$ \mathcal{H}_T = \bigcup_{x\in M} T_x^\perp \,, \qquad {\rm where} \qquad
T_x^\perp = \{ \mathbf{Y}_x \in T_x\mathcal{M} | T_x(\mathbf{Y}_x,\_) = 0 \}  $$
It is a differential system of constant rank $4-m$ (see \cite{Godbillon}, sections V.3 and V.4)  and the {\em associated Pfaff system} is $\mathcal{P}_T = \mathcal{H}_T^\perp \subset \Lambda^1\mathcal{M}$.

$T$ is said to be {\em holonomous} if its associated Pfaff system is integrable and, in such a case, local charts $(y^1, \ldots y^4)$ exist such that $ \phi^\alpha = a^\alpha_\beta(y^b) \,dy^\beta  \,$ ---see ref. \cite{Godbillon}, Lemma V.4.10. 
In terms of these coordinates, 
\begin{equation} \label{MC2.200}
T = T_{\alpha\beta}(y^b) \,dy^\alpha \otimes dy^\beta    \qquad {\rm and} \qquad \mathbf{X} = X^\alpha \frac{\partial}{\partial y^\alpha} + X^A \partial_A \,, \qquad \partial_A:= \frac{\partial}{\partial y^A}  \,,
\end{equation}
where $T_{\alpha\beta}(y^b) = \eta_{\mu\nu} a^\mu_\alpha (y^b)\,a^\nu_\beta (y^b)$.

%%%%%%%%%%%%%%%%%%%%%%%%%%%%%%%%%%%%%%%%%%%%%%%%%%
\section{Collineations of a rank 3 tensor \label{S2}}
If ${\rm rank} T = 3$, it is obvious that $T$ is holonomous and local charts exist such that the expressions (\ref{MC2.200}) hold.
We write $\mathbf{N}:= \partial_4$ and $T= T_{\alpha\beta} dx^\alpha\otimes dx^\beta$, with $\det T_{\alpha\beta} \neq 0$ and then decompose the collineation field as
$$\mathbf{X} = \mathbf{Z} + f\,\mathbf{N} \,\qquad \mbox{where $f$ is a function} $$
and $\mathbf{Z} = Z^\alpha \partial_\alpha $ is tangent to the submanifolds $y^4=\,$constant. 

It is obvious that $T( \mathbf{N},\,\_)=0$, which implies that $\mathcal{L}_{f\mathbf{N}} T = f\,\mathcal{L}_{\mathbf{N}} T $ and therefore, equation (\ref{e1}) amounts to  
\begin{equation}\label{MC3.1}
\mathcal{L}_{\mathbf{Z}} T + f \mathcal{L}_{\mathbf{N}} T = 0 
\end{equation}
The projections of this  over $\mathbf{N}$ and over the submanifolds $y^4=\,$constant respectively  yield 
\begin{equation}\label{MC3.2}
[\mathbf{N},\mathbf{Z}] = 0 \qquad {\rm and } \qquad \mathcal{L}_{\mathbf{Z}} T + 2 f K = 0  
\end{equation}
where $K := \frac12\,\partial_4 T_{\alpha\beta} dy^\alpha \otimes dy^\beta $ or, in components,
\begin{equation}\label{MC3.3}
 \partial_4 Z^\alpha = 0 \qquad {\rm and } \qquad \nabla_{(\alpha} Z_{\beta)} +  f K_{\alpha\beta}= 0
\end{equation}
where $\nabla$ is the Levi-Civita connection for the non-degenerate metric $T_{\alpha\beta}$ on the hypersurfaces $y^4={\rm constant}$.
The second of these equations looks like a non-homogeneous Killing equation (parametrized with $y^4$\,) and  the question is: does it admit solutions $Z^\alpha$ that do not depend on $y^4$ for some appropriate $f$?  

If $K_{\alpha\beta}=0$, i. e.   $\mathcal{L}_{\mathbf{N}} T = 0$, the answer is obvious because it reduces to a Killing equation in 3 dimensions. The collineation field is then $\mathbf{X}= \mathbf{Z} + f\,\mathbf{N}$, where $f$ is arbitrary and $\mathbf{Z}$ is a Killing vector for the non-degenerate metric $T$ in each submanifold $y^4$ constant. 

If $K_{\alpha\beta}\neq 0$, things are not so simple. Let us start from equations (\ref{MC3.3}) ---that are equivalent to (\ref{MC3.2})---, the second of them implies that
\begin{equation}\label{MC3.4}
\nabla_{\alpha} Z_{\beta} = \Omega_{\alpha\beta} -  f K_{\alpha\beta} \qquad {\rm with} \qquad  \Omega_{\alpha\beta}+ \Omega_{\beta\alpha}=0
\end{equation}
The integrability conditions imply new equations on $\Omega_{\alpha\beta}$ and $f$. These can be derived by means of the Lie derivative of a connection ---see ref. \cite{Yano}, section I.4\,. We have that
\begin{equation}\label{MC3.4a}
\mathcal{L}_\mathbf{Z} \nabla_\mu T_{\kappa\lambda} - \nabla_\mu \mathcal{L}_\mathbf{Z}  T_{\kappa\lambda} =
- \left( \mathcal{L}_\mathbf{Z} \Gamma^\alpha_{\mu\kappa} \right)\, T_{\alpha\lambda} -
\left( \mathcal{L}_\mathbf{Z} \Gamma^\alpha_{\mu\lambda}\right)\, T_{\kappa\alpha}
\end{equation}
with 
\begin{equation}\label{MC3.6}
\mathcal{L}_\mathbf{Z} \Gamma^\alpha_{\mu\kappa} = \nabla_\mu \nabla_\kappa Z^\alpha - Z^\rho R_{\rho\mu\kappa}^{\hspace*{1em}\;\alpha}
\end{equation}
Then, as for the Levi-Civita connection $\nabla_\mu T_{\kappa\lambda} = 0$, from (\ref{MC3.4a}) it easily follows that also
\begin{equation}\label{MC3.5}
\mathcal{L}_\mathbf{Z} \Gamma^\alpha_{\mu\kappa} = \frac12\, T^{\alpha\lambda}\,\left( \nabla_\mu \mathcal{L}_\mathbf{Z}  T_{\kappa\lambda} + \nabla_\kappa \mathcal{L}_\mathbf{Z}  T_{\mu\lambda} - \nabla_\lambda \mathcal{L}_\mathbf{Z}  T_{\mu\kappa}
\right)
\end{equation}
where $T^{\alpha\lambda} T_{\lambda\mu} = \delta^\alpha_\mu$.

Combining then equations (\ref{MC3.3}), (\ref{MC3.5}), (\ref{MC3.6}) and including (\ref{MC3.4}), we arrive at:
\begin{equation}\label{MC3.7}
\nabla_\mu \Omega_{\kappa\lambda}  = Z^\rho R_{\rho\mu\kappa\lambda} + \nabla_\lambda \left( f K_{\mu\kappa}\right) - \nabla_\kappa \left( f K_{\mu\lambda}\right)
\end{equation}

Furthermore, for the connection we have that\footnote{Notice that the definition of the Riemann tensor in \cite {Yano} and the common definition in other references cited in this text differ in one sign, which we have included}  ---see \cite{Yano}, equation (I.4.14)---
$$  \nabla_\nu \mathcal{L}_\mathbf{Z} \Gamma^\kappa_{\mu\lambda} - \nabla_\mu \mathcal{L}_\mathbf{Z} \Gamma^\kappa_{\nu\lambda} = - \mathcal{L}_\mathbf{Z} R_{\nu\mu\lambda}^{\hspace*{1em}\;\kappa}  $$
which, including (\ref{MC3.5}) yields
$$ \mathcal{L}_\mathbf{Z} R_{\nu\mu\lambda\kappa} = - 2 f R^\rho_{\;[\kappa \nu\mu} K_{\lambda]\rho} - 2 \nabla_{[\nu} \nabla_\kappa \left( fK_{\mu]\lambda} \right) + 2 \nabla_{[\nu} \nabla_\lambda \left( fK_{\mu]\kappa} \right)   \,.$$
As $\nu, \mu, \ldots = 1,2,3$ and both sides of this equation have the same symmetries as a Riemann tensor, it is equivalent to one of its traces, namely:
\begin{equation}\label{MC3.8}
\mathcal{L}_\mathbf{Z} R_{\nu\lambda} =  2 f \,\left( K^{\mu\kappa}R_{\nu\mu\lambda\kappa} - R^\alpha_{\;(\nu} K_{\lambda)\alpha} \right) + \nabla^\alpha\nabla_\alpha \left( f K_{\nu\lambda} \right) +  \nabla_{\nu} \nabla_\lambda \left( f K^\mu_{\;\mu} \right) - 2  \nabla_{(\nu} \nabla_\alpha \left( f K^\alpha_{\;\lambda)} \right) 
\end{equation}

Then, using a relation analogous to (\ref{MC3.4a}) for tensors ---see \cite{Yano}, equation (I.4.9)--- we obtain the hierarchy of relations 
$$ \mathcal{L}_\mathbf{Z} \nabla_{\alpha_1} \ldots \nabla_{\alpha_n} R_{\nu\lambda} = H_{\alpha_1 \ldots \alpha_n \nu\lambda} \left( f, f_{\|\alpha} \ldots f_{\|\beta_1\ldots \beta_{n+2} }, \ldots \right) \qquad {\rm where} \qquad f_{\|\alpha \ldots \beta }:= \nabla_\beta \ldots \nabla_\alpha f \,,$$ 
similarly as in the theory of Killing vectors (see \cite{Cramer}, chapter 8).

What has been done so far amounts to analysing the integrability conditions derived from the commutation of $\nabla_\mu$ and $\nabla_\nu$. Let us now turn to the commutation of $\nabla_\nu$ and $\mathbf{N} =\partial_4$. 

The first of equations (\ref{MC3.2}) implies that $\left[\mathcal{L}_\mathbf{Z},\mathcal{L}_\mathbf{N} \right]=0$ which, applied to the second equation (\ref{MC3.2}) yields $ \mathcal{L}_\mathbf{Z} K_{\alpha\beta} + \partial_4 \left( f K_{\alpha\beta} \right) = 0  \,$
or, rising the index with $T^{\mu\alpha}$, 
\begin{equation}\label{MC3.9}
\mathcal{L}_\mathbf{Z} K^\mu_{\;\beta} + \dot f K^\mu_{\;\beta} + f \dot K^\mu_{\;\beta} = 0 
\end{equation}
where $\dot f := \partial_4 f$ and $\dot K^\mu_{\;\beta} := \partial_4 K^\mu_{\;\beta}$. % and, by iterated application of $\mathcal{L}_\mathbf{N}$, we also obtain that
%\begin{equation}\label{MC3.9a}
%\mathcal{L}_\mathbf{Z} K^{(n)\mu}_{\hspace*{1em}\;\beta} + \partial_4^{n+1} \left( f K^\mu_{\;\beta} \right) = 0 
%\end{equation}
%where $K^{(n)\mu}_{\hspace*{1em}\;\beta} := \partial_4^n  K^\mu_{\;\beta}$.

Similarly, by applying the commutation relation $\left[\mathcal{L}_\mathbf{Z},\mathcal{L}_\mathbf{N} \right]=0$ to equations (\ref{MC3.5}), (\ref{MC3.8}) and to the whole above hierarchy, we should obtain further algebraic relations connecting $Z^\lambda$, $\Omega_{\mu\nu}$, $f$, $\dot f$, $f_{|\mu}$, $\dot f_{|\mu}$, $f_{\|\mu\nu}$, etc, which we shall not write explicitely.

Also, applying $\partial_4$ to (\ref{MC3.4}) and including  (\ref{MC3.3}) we have that
$$\partial_4 \Omega_\mu^{\;\nu} = \partial_4\left(f K^\nu_{\;\mu} \right) + \dot\Gamma^\nu_{\mu\alpha} Z^\alpha $$
where $ \dot\Gamma^\nu_{\mu\alpha}:= \partial_4\Gamma^\nu_{\mu\alpha} = \nabla_\mu K^\nu_{\;\alpha} + \nabla_\alpha K^\nu_{\;\mu} - \nabla^\nu K_{\mu\alpha} $ and, using (\ref{MC3.9}) it follows that
\begin{equation}\label{MC3.10}
\partial_4 \Omega_{\mu\nu} = 2 \Omega_{\lambda[\nu} K^\lambda_{\;\mu]} + 2 Z^\lambda \nabla_{[\mu} K_{\nu]\lambda}
\end{equation}

Turning now back to equation (\ref{MC3.9}), unless $\dot K^\mu_{\;\beta} \propto K^\mu_{\;\beta}$, it permits to derive $f$ as  a linear function of $Z^\lambda$ and $\Omega_{\mu\nu}$. Indeed, if $\dot K^\mu_{\;\beta}$ is not proportional to $K^\mu_{\;\beta}$, it exists $M^\alpha_{\;\mu}$ such that 
$ M^\alpha_{\;\mu} K^\mu_{\;\alpha} = 0$  and $M^\alpha_{\;\mu}  \dot K^\mu_{\;\alpha} =1$; therefore 
$$ f = - M^\alpha_{\;\mu} \,\mathcal{L}_\mathbf{Z} K^\mu_{\;\alpha} = -  M^\alpha_{\;\mu} \,\left(Z^\beta \nabla_\beta K^\mu_{\;\alpha} + 
K^\mu_{\;\beta}   \Omega_\alpha^{\;\beta} - K^\beta_{\;\alpha} \Omega_\beta^{\;\mu}  \right)$$
(If there are more than one independent matrix $M^\alpha_{\;\beta}$ fulfilling the above trace equalities, it will result in constraints connecting $Z^\alpha$ and $\Omega_{\mu\beta}$.)

Substituting then this $f$ in equations (\ref{MC3.3}), (\ref{MC3.7}) and (\ref{MC3.10}) we obtain a closed partial differential system on $Z^\alpha$ and $\Omega_{\mu\nu}$ (with no extra functions). If it is integrable, each solution is parametrized by six real numbers, namely $Z^\alpha(0)$ and $\Omega_{\mu\nu}(0)$, i. e. the values of the unknowns at one point. The above mentioned hierarchy of integrability conditions then act as constraints on these parameters and the number of dimensions of the collineation algebra $\mathcal{C}_T$ is at most six. 

If, on the contrary, $\dot K^\mu_{\;\beta} = b  K^\mu_{\;\beta}$, then equation (\ref{MC3.9}) implies that 
$$ \mathcal{L}_\mathbf{Z} K^\mu_{\;\beta} + (\dot f + b f ) \,K^\mu_{\;\beta} =0 \,, \qquad \mbox{for some } b\,, 
$$
which allows to derive $\dot f $ as a linear function of $Z^\alpha$, $\Omega_{\mu\nu}$ and $f$. Indeed, as $K^\mu_{\;\beta} \neq 0$, it exists $N^\alpha_{\;\beta}$ such that $ N^\alpha_{\;\mu} K^\mu_{\;\alpha} = 1$; therefore
\begin{equation}\label{MC3.12}
\dot f + bf = - N^\alpha_{\;\mu} \,\mathcal{L}_\mathbf{Z} K^\mu_{\;\alpha} = -  M^\alpha_{\;\mu} \,\left(Z^\beta \nabla_\beta K^\mu_{\;\alpha} + 
K^\mu_{\;\beta}   \Omega_\alpha^{\;\beta} - K^\beta_{\;\alpha} \Omega_\beta^{\;\mu}  \right)
\end{equation} 

Now, applying $\partial_4$ to both sides of equation (\ref{MC3.7}), we obtain that (see the Appendix)
\begin{equation}\label{MC3.14}
f_{\|\alpha} \left(K^{\lambda\alpha} K_{\mu\kappa} - K_{\mu\beta} K^\beta_{\;\kappa} T^{\alpha\lambda} \right) = W^\lambda_{\mu\kappa} 
\end{equation}
where $W^\lambda_{\mu\kappa}$ is a linear function of $Z^\alpha$ and $\Omega_{\mu\nu}$. 
In some cases this permits to derive $f_{\|\alpha}$ as a linear function of $Z^\alpha$, $\Omega_{\mu\beta}$ and $f$. Indeed,  $ K^{\lambda\alpha} K_{\mu\kappa} - K_{\mu\beta} K^\beta_{\;\kappa} T^{\alpha\lambda} $ can be seen as a linear map from the 4-dimensional space $f_{\|\alpha}$ into the $4\times10$ space $W^\lambda_{\mu\kappa} $ and it can be inverted whenever (a) it is injective, which only fails to happen if $K^\alpha_{\;\mu} K^\mu_{\;\beta} = 0$ or $K^\alpha_{\;\mu} \propto \delta^\alpha_{\;\mu}$, and (b) the right hand side $W^\lambda_{\mu\kappa}$ fulfills some conditions, i. e. some linear constraints on $Z^\alpha$, $\Omega_{\mu\beta}$ and $f$.  

This $f_{\|\alpha}$, written as a linear function of $Z^\alpha$, $\Omega_{\mu\beta}$ and $f$, together with (\ref{MC3.3}), (\ref{MC3.4}), (\ref{MC3.7}) and (\ref{MC3.12}), yields a partial differential system on the variables $Z^\alpha$, $\Omega_{\mu\nu}$ and $f$. If it is integrable, each solution is parametrized by seven real numbers, namely $Z^\alpha(0)$, $\Omega_{\mu\nu}(0)$  and $f(0)$, the values of the variables at a point. The above mentioned hierarchy of integrability conditions are to be taken as constraints on these parameters and the number of dimensions of the collineation algebra $\mathcal{C}_T$ is at most seven. 

%%%%%%%%%%%%%%%%%%%%%%%%%%%%%%%%%%%%%%%%%%%%%%%%%%
\section{Collineations of a rank 1 tensor  \label{S3.1}}
If rank$\,T = 1$, it can be written locally as $T= \pm \,\phi \otimes \phi \,$, $\, \phi\in\Lambda^1\mathcal{M}\, $, and
the collineation condition $\mathcal{L}_{\mathbf{X}} T =0$ is equivalent to $\mathcal{L}_{\mathbf{X}} \phi =0$,
which means that, locally, a function $f$ exists such that 
\begin{equation} \label{e6}
\mathbf{(a)}\qquad  i_\mathbf{X} \phi = f \qquad {\rm and} \quad
\mathbf{(b)}\qquad  i_\mathbf{X} d\phi = - df
\end{equation}
which is a linear system on $\mathbf{X}$ whose compatibility depends on $f$ and on the ranks of the differential forms $\phi$ and $d\phi$.
The general solution $\mathcal{I}_f$  can be written as
$$ \mathcal{I}_f = \mathbf{X}_f + \mathcal{I}_0 $$
where $\mathbf{X}_f$ is a particular solution and $\mathcal{I}_0 = \{\mathbf{Y} |\,i_\mathbf{Y}\phi = i_\mathbf{Y} d\phi = 0\} $ is the general solution of the homogeneous system.

To study the compatibility of (\ref{e6}), we invoke the following corollary of Darboux theorem --- see \cite{Godbillon}, Theorem VI.4.1--- 
\begin{theorem}  \label{T1}
Given $\phi \in\Lambda^1\mathcal(M)$, they exist a canonical coordinate system $p_1, p_2, q^1 , q^2$ and a function $\psi$, such that
\begin{equation}  \label{e7}
\phi =d\psi + e_1 p_1 dq^1 + e_2 p_2 dq^2\,, \qquad {\rm with} \qquad e_1\geq e_2 \qquad e_1,e_2 = 0,1
\end{equation}
\end{theorem}

A remark on notation is appropriate: hereon a stroke means partial derivative, so $v_{|a}:=\partial_a v:=\partial v/\partial x^a$, $a=1, \ldots 4$; particularly in canonical coordinates $(q^i, p_j)$, we shall write 
$$ v_{|i}:=\partial_i v:= \frac{\partial v}{\partial q^i} \qquad {\rm and} \qquad v^{|j}:=\partial^j v:= \frac{\partial v}{\partial p_j}\,, \qquad i,j = 1, 2 $$
Writing now $\mathbf{X}$ and $df$ in canonical coordinates, we have
$$ \mathbf{X}= X^i\,\partial_i + X_i\,\partial^i \,, \qquad  \qquad df = f_{|i} dq^i + f^{|i} dp_i $$
and (\ref{e6}.$\mathbf{b}$) amounts to
\begin{equation} \label{e8}
-e_i X_i = f_{|i} \,, \qquad e_i X^i = f^{|i}
\end{equation}
Then, substituting this and (\ref{e7}) into (\ref{e6}.$\mathbf{a}$), we obtain that the latter amounts to
\begin{equation} \label{e9}
X^i \phi_i+ X_i \phi^i = f   \,, \qquad {\rm with} \qquad
\phi_i:=\psi_{|i}+ e_i p_i \qquad {\rm and} \qquad \phi^i:=\psi^{|i}
\end{equation}

According to the values of $e_1$ and $e_2$, different cases are possible, which we shall analyse separately:
\begin{description}
\item[{[1.nd]} ] \underline{$(d\phi)^2\neq 0$}. Then $\Omega:=d\phi$ is a symplectic form and $e_1=e_2=1$.\\[.5ex]
     In this case the class of the differential form $\phi$ is 4 ---see \cite{Godbillon}, Section VI.1.3--- and Darboux theorem states more precisely that canonical local charts exist such that $\psi=0$, that is $\phi = p_1 dq^1+p_2 dq^2$. Equation (\ref{e8}) then implies that 
     $$  X_i = -  f_i \qquad {\rm and} \qquad  X^i = f^i  \,,\qquad {\rm or} \qquad 
     \mathbf{X} =- \{f,\_ \} \,, $$ 
     where $\{\;,\;\}$ is the Poisson bracket with elementary Poisson brackets: $\{q^i,p_j\}=\delta^i_j$, $i,j=1,2$. Substituting this in equation (\ref{e9}), it becomes 
     \begin{equation} \label{e9a}
     \sum_{i=1}^2 p_i f^{|i} = f 
     \end{equation}
     which, by Euler theorem, means that $f(q^i,p_j)$ is an homogeneous function of the first degree in the variables $p_j$.      The general collineation field is thus
     $ \mathbf{X} = - \{f,\_ \} \;$, where $f\in \Lambda^0\mathcal{M}$ is a solution of (\ref{e9a}).

\item[{[1.d]}] \underline{$(d\phi)^2=0$} but $d\phi \wedge \phi \neq 0$ which, including Darboux Theorem, implies that $\phi=d\psi + p_1\, dq^1$ with $d\psi\wedge dp_1\wedge dq^1\neq 0$. In this case, local charts of canonical coordinates exist such that $d\phi = d q^2 + dp_1\wedge dq^1\,$, i. e.  $e_2=0$ and $e_1=1$. Combining then equations (\ref{e9}) and (\ref{e8}), we obtain that 
     \begin{equation} \label{e9b}
    X_1 = -f_{|1} \,, \qquad X^1 = f^{|1} \,, \qquad f=f(p_1,q^1)  \qquad {\rm and} \qquad      X^2= f-p_1\,f^{|1} 
     \end{equation}
     The component  $X_2$ is not determined and the general collineation field is
\begin{equation} \label{e9c}
 \mathbf{X} =  f^{|1} \,\partial_1 - f_{|1} \,\partial^1 + (f-p_1\,f^{|1})\,\partial_2 + X_2\,\partial^2
\end{equation}
where $f(p_1,q^1)$ and $ X_2(p_i,q^j)$ are arbitrary functions of their respective variables.
\item[{[1.d.h]}] \underline{$(d\phi)^2=d\phi \wedge \phi=0$}  but $d\phi \neq 0$. In this case $\phi$ is integrable and a local chart exists such that $\phi = p_1 dq^1$ combining then equations (\ref{e9}) and (\ref{e9b}), we obtain that 
$$ f- p_1 f^{|1} = 0 \qquad \mbox{or, equivalently} \qquad f = p_1 \,F(q^1)  $$
There is no constraint on the components $X^2$ and $X_2$ and the general collineation field,
\begin{equation} \label{e9d}
 \mathbf{X} = F(q^1)\,\partial_1 - p_1 F^\prime(q^1) \,\partial^1 + X^2 \,\partial_2 +X_2 \,\partial^2 \,, 
\end{equation}
which contains three arbitrary functions, namely $ F(q^1)$, $X^2(p_i,q^j)$ and $X_2(p_i,q^j)$.
\item[{[1.d.0]}] \underline{$d\phi=0$}. Then, it exists $\psi$ such that $\phi = d\psi$, i.e. (locally) an exact differential. 
The equation $i_{\mathbf{X}}d\phi = -df$ implies that  $f= C$, constant, with no further restrictions on $\mathbf{X}$. 
The  other equation, $i_{\mathbf{X}}\phi = C$, then reads $\mathbf{X}\psi = C$ and, in a local chart $\{x^a\}_{a=1\ldots 4}$, such that $x^1=\psi$, the general collineation field is
 \begin{equation} \label{e9f}
  \mathbf{X} = C\,\frac{\partial \;\;}{\partial x^1} + \sum_{\nu=2}^4 X^\nu \,\frac{\partial \;\;}{\partial x^\nu} 
  \end{equation}
with $X^\nu(x^a)$  arbitrary.
\end{description}

\section{Collineations of a rank 2 tensor   \label{S3}}
Now $T=\eta_{\alpha\beta}\,\phi^\alpha\otimes\phi^\beta$, with $\eta_{\alpha\beta}= {\rm diag}\,(1,\sigma)\,$, $\sigma=\pm 1$. 
In what follows it will be helpful to consider the 2-forms $d\phi^\alpha$ and the exterior products
\begin{equation}  \label{MC2.2}
\Upsilon^\alpha := d\phi^\alpha \wedge \phi^1 \wedge \phi^2 \qquad {\rm and} \qquad 
\Sigma^{\alpha\beta} := d\phi^\alpha \wedge d\phi^\beta \,.
\end{equation}

Under a $T$-rotation (\ref{MC2.1}) we have that 
$$d\tilde\phi^\alpha = d R^\alpha_{\;\beta} \wedge \phi^\beta + R^\alpha_{\;\beta} \,d\phi^\beta \qquad {\rm and} \qquad  \tilde\phi^1 \wedge \tilde\phi^2 = \det(R^\alpha_{\;\beta})\,  \phi^1 \wedge \phi^2 \,,$$
and, as  $\det(R^\alpha_{\;\beta}) = \pm 1$, it follows that 
\begin{equation}  \label{MC2.3}
\tilde\Upsilon^\alpha := \pm \,R^\alpha_{\;\beta} \Upsilon^\beta %\qquad {\rm and} \qquad  
\end{equation}
Now, let $\Omega \in \Lambda^4\mathcal{M}$ be a volume tensor ($\Omega \neq 0$) and let us define $l^\alpha$ by  $\Upsilon^\alpha = l^\alpha \Omega$. The relation (\ref{MC2.3}) implies that $ \tilde l^\alpha := \pm R^\alpha_{\;\beta} l^\beta $
and, as $R^\alpha_{\;\beta}$ is a $T$-rotation, we have that
\begin{center} 
 $\eta_{\alpha\beta} l^\alpha l^\beta $ is invariant by $T$-rotations
\end{center}

As a consequence, unless $\eta_{\alpha\beta} = {\rm diag}(1, -1) $ and $\Upsilon^1 = \Upsilon^2$, we can allways perform a $T$-rotation such that one of the exterior products $\Upsilon^\alpha$ vanishes. (We can label the 1-forms $\phi^\beta$ so that this is $\Upsilon^1$.) 
Therefore, $T$ can be classified in one of the following types:
\begin{center}
\begin{tabular}{|c|cl|}   \hline
{\bf 2.I} & $\Upsilon^1= 0 \,, \quad \Upsilon^2 \neq 0$ & \begin{tabular}{|c|l}
                                         {\bf a} & $\Sigma^{11}\neq 0$ \\ \hline
                                         {\bf b} & $\Sigma^{11}= 0 $ % \\ \hline 
                                         \end{tabular} \\ \hline
{\bf 2.N} & $\Upsilon^1= \Upsilon^2\neq 0 $ &    \\  \hline
{\bf 2.H} & $\Upsilon^1= \Upsilon^2 = 0$   &    \\  \hline
\end{tabular}
\end{center}
[Notice that Type 2.N only occurs if $\eta_{\alpha\beta} = {\rm diag}(1, -1) $\,].

%%%%%%%%%%%%%%%%%%%%%%%%%%%%%%%%%%%%%%%%%%%%%%%%%%%%%%%%%%%%%%%%
\subsection{Type 2.I.a  \label{S3.2.a}}
\begin{proposition}  \label{p.MC2}
If $\Sigma^{11} \neq 0$, $\Upsilon^1 = 0$ and $\Upsilon^2\neq 0$, then two differential forms $\psi_\alpha \in \Lambda^1\mathcal{M}$, $\alpha = 1,\,2$, exist such that 
\begin{equation} \label{MC2.6}
d\phi^1=\psi_\alpha \wedge \phi^\alpha \,, \qquad   d\phi^2= r \,d\phi^1 + \frac{s-r^2}{2 l}\,\phi^1 \wedge \phi^2 -2 l \,\psi_1 \wedge \psi_2 \,.
\end{equation}
where
\begin{equation} \label{MC2.6a}
\Sigma^{12}= r \Sigma^{11} \, ,\qquad  \Sigma^{22}= s \Sigma^{11} \qquad {\rm and} \qquad \Upsilon^2 = l \Sigma^{11} \,, \qquad l \neq 0 
\end{equation} 
The differential forms $\psi_\alpha$ are uniquely determined and $\{\phi^\alpha, \,\psi_\beta\}_{\alpha,\beta=1,2}$ is the {\em canonical base} for the tensor $T$. 
\end{proposition}

\paragraph{Proof:} The first expression in (\ref{MC2.6}) follows immediately from $\Upsilon^1=0$ ---see ref. \cite{Godbillon}, Chapter V, Proposition 4.12\,. Then the fact that $\Sigma^{11}  = - 2 \phi^1 \wedge \phi^2 \wedge \psi_1 \wedge \psi_2 \neq 0 $ implies that $\{\phi^\alpha, \,\psi_\beta\}_{\alpha,\beta=1,2}$ are independent. 

The 1-forms $\psi_\alpha$ are determined apart from the gauge freedom:
\begin{equation} \label{MC2.7}
\psi^\prime_\beta = \psi_\beta + B_{\alpha\beta} \phi^\alpha \qquad {\rm with} \qquad B_{\alpha\beta} = B_{\beta\alpha}
\end{equation}

We now write $d\phi^2= P^\alpha_{\;\beta} \,\psi_\alpha\wedge \phi^\beta + a\,\phi^1 \wedge \phi^2 + m \,\psi_1 \wedge \psi_2$ and, including that 
$\Sigma^{11} = - 2 \phi^1 \wedge \phi^2 \wedge \psi_1 \wedge \psi_2$, we have from (\ref{MC2.6a}) that 
$$ m = - 2 l\,, \qquad 2 r = P^\alpha_{\;\alpha} \,, \qquad s = 2 a l + \det(P^\alpha_{\;\beta}) $$

Under the gauge transformations (\ref{MC2.7}) the components of $d\phi^2$ change according to:
$$ P^{\prime \alpha}_{\;\;\beta} = P^{\alpha}_{\;\beta} - m \,B_{\beta\nu} \epsilon^{\alpha\nu} \,, \qquad 
a^\prime = a - B_{\nu\alpha}\epsilon^{\nu\beta} P^{\alpha}_{\;\beta} + m \,B_{\beta 1} B_{\alpha 2} \epsilon^{\beta\alpha} $$
whereas $m = - 2 l$ and $P^{\alpha}_{\;\alpha} = 2 r$ are gauge invariant. We can therefore choose the gauge matrix $B_{\beta\alpha}$ so that the traceless part $P^{\prime \alpha}_{\;\;\beta}$ vanishes. That is, the base 1-forms $\psi_\alpha$  can be chosen so that $P^{\alpha}_{\;\beta}= r\,\delta^{\alpha}_{\beta}$ and the second and third expressions in (\ref{MC2.6}) follow immediately. 

Notice also that the above choices exhaust the gauge freedom. \hfill $\Box$

As rank$\,T = 2$, equation (\ref{e4}) reads
\begin{equation} \label{MC2.8}  %13}
\mathcal{L}_{\mathbf{X}} \phi^\alpha = b \,D^\alpha_{\;\beta} \phi^\beta \,, \qquad {\rm where} \qquad D^\alpha_{\;\beta} := \eta^{\alpha\nu} \epsilon_{\nu\beta} = \left(\begin{array}{cc}
                            0 & 1 \\ -\sigma & 0 
                            \end{array}\right) \,,
\end{equation}
$\epsilon_{\nu\beta}  = - \epsilon_{\beta\nu} \,$, $\epsilon_{12} =1$ and $b$ is a function. Therefore it follows that 
$$ \mathcal{L}_{\mathbf{X}} \left(\phi^1 \wedge \phi^2\right) =0 \qquad {\rm and}  \qquad \mathcal{L}_{\mathbf{X}} d\phi^\alpha = D^\alpha_{\;\beta}\,db \wedge \phi^\beta + b\,D^\alpha_{\;\beta}\,d\phi^\beta  $$ 
Then for  $\Upsilon^\alpha$ we have that  
$\, \mathcal{L}_{\mathbf{X}}\Upsilon^1  = b\,\Upsilon^2 \, $ and, as $\Upsilon^1 =0$ and $\Upsilon^2 \neq 0$, it follows that $b=0$ which, substituted in (\ref{MC2.8}) yields 
\begin{equation} \label{MC2.9}   %  14}
\mathcal{L}_{\mathbf{X}} \phi^\alpha = 0 
\end{equation}

On their turn, these equations imply that $\mathcal{L}_{\mathbf{X}} d\phi^\alpha = 0 $ which lead to $\,\mathcal{L}_{\mathbf{X}}\Sigma^{\alpha\beta} = \mathcal{L}_{\mathbf{X}} \Upsilon^\alpha =0\,$ and
\begin{equation} \label{MC2.10}  
\mathbf{X} l = \mathbf{X} r = \mathbf{X} s = 0
\end{equation}

Including this and equation (\ref{MC2.6}), $ \mathcal{L}_{\mathbf{X}} d\phi^\alpha = 0$ implies that 
$$ \mathcal{L}_{\mathbf{X}} \psi_\alpha \wedge \phi^\alpha = 0 \qquad {\rm and}  \qquad
\mathcal{L}_{\mathbf{X}} \psi_1 \wedge \psi_2 + \psi_1  \wedge \mathcal{L}_{\mathbf{X}} \psi_2 = 0  \,, $$
whence it easily follows that 
\begin{equation} \label{MC2.11}
\mathcal{L}_{\mathbf{X}} \psi_\alpha = 0 
\end{equation}

Summarizing, if $T$ is type {\bf 2.I.a}, first we find the canonical base $\{\phi^a\}_{a=1\ldots 4}$, where $\phi^3 := \psi_1$ and \\
$\phi^4 := \psi_2$, and its dual base $\{\mathbf{Y}_a\}_{a=1\ldots 4}$. Then the collineation equations supplemented with their integrability conditions amount to $\mathcal{L}_{\mathbf{X}} \phi^a = 0 $ or, writing 
$\mathbf{X}= X^a \mathbf{Y}_a \,$ and $ \, d\phi^a = -\frac12\, C^a_{bc} \phi^b \wedge \phi^c \,$, 
\begin{equation} \label{MC2.12}
 d X^a - X^b C^a_{bc} \phi^c = 0
\end{equation}
%(Part of the commutation coefficients $C^a_{bc}$  can be derived from Proposition \ref{p.MC2} and are given in Appendix A.) 
If this partial differential system is integrable, each solution is parametrized by the values $X^b_0$ at one point. Therefore the dimension of the collineation algebra for type {\bf 2.I.a} tensors is at most 4.

The integrability conditions of (\ref{MC2.12})  put some further constraints on the parameters $X^b_0$. These integrability conditions are obtained by taking the exterior derivative and  read   $\mathcal{L}_{\mathbf{X}} d\phi^a = 0$ or, in terms of the coefficients $C^a_{bc}$, 
\begin{equation} \label{MC2.12a}
 \mathbf{X} C^a_{bc}  = 0
\end{equation}
Locally this amounts to $\left[\mathbf{X} C^a_{bc}\right]_0 = 0$, which is an algebraic constraint on $X^b_0$, plus $d\left(\mathbf{X} C^a_{bc}\right) = 0$. 

Using the fact that $d$ and $\mathcal{L}_\mathbf{X}$ commute, the latter is equivalent to:
$$ \mathbf{X} C^a_{bc|e} = 0 \,,\qquad {\rm where} \qquad  C^a_{bc|e}:= \mathbf{Y}_e C^a_{bc}  $$
Iterating this procedure, we obtain that (\ref{MC2.12a}) implies that
\begin{equation} \label{MC2.12b}
  \left.C^a_{bc|e_1 \ldots e_n h}\right|_0 X^{h}_0 = 0 \,, \qquad n \in\mathbb{N}
\end{equation}
which is an infinite homogeneous linear system on the parameters $X^b_0$. Provided that its rank is not greater than 4, the codimension of the collineation algebra for type {\bf 2.I.a} tensors is precisely this rank, otherwise $T$ admits no collineation fields. 

%%%%%%%%%%%%%%%%%%%%%%%%%%%%%%%%%%%%%%%%%%%%%%%%%%%%%%%%%%%%%%%%
\subsection{Type 2.I.b   \label{S3.2.b}}
\begin{proposition}  \label{p.MC3}
If $\Upsilon^2 \neq 0$, $\Upsilon^1=0$ and $\Sigma^{11} = 0$, then two differential forms $\phi^A \in \Lambda^1\mathcal{M}$, $A=3,\,4$, exist such that 
\begin{equation} \label{MC2.16}
d\phi^2= \frac{s}{2}\, \phi^1 \wedge \phi^2 + \phi^3 \wedge \phi^4\,, \qquad   d\phi^1= r \,\phi^1 \wedge \phi^2 + v_\alpha\,\phi^\alpha \wedge \phi^3 \,.
\end{equation}
where either $v_\alpha=(0,0)$, $v_\alpha= (1,v)$ or $v_\alpha = (v, 1)$. Besides $ \Sigma^{12}= r \Upsilon^2 $  and $\, \Sigma^{22}= s \Upsilon^2 \,$. 

The differential form $\phi^4$ is determined up to the gauge transformation, $\phi^{\prime 4} = \phi^4 + m \phi^3$, where $m$ is an arbitrary function.
\end{proposition}

\paragraph{Proof:} Consider $\Theta:=\displaystyle{d\phi^2 - \frac{s}{2}\,\phi^1 \wedge \phi^2}$. By the hypothesis, $\Theta \wedge \Theta =0$, which implies that $\Theta$ is simple and $\phi^A$, $A=3,4$, exist such that $\Theta = \phi^3 \wedge \phi^4 $. These $\phi^A$ present an obvious $SL(2)$ gauge freedom.

As $\Upsilon^2 \neq 0$, the four 1-forms $\{\phi^a \}_{a=1\ldots 4}$ are independent. Writing then $d\phi^1$ in this base and including that $\Sigma^{12}= r \Upsilon^2$, it immediately follows that $d\phi^1= r \,\phi^1 \wedge \phi^2 + P_{\alpha B}\,\phi^\alpha \wedge \phi^B$ and, as $\Sigma^{11}=0$, $\det\,\left( P_{\alpha B}\right)_{\alpha=1,2; A=3,4} =0$; therefore $v_\alpha$ and $P_B$ exist such that $P_{\alpha B} = v_\alpha P_B$. Then the gauge freedom in the definition of $\phi^B$ can be used to make $P_B = \delta^3_B$ and, if $v_\alpha \neq 0$, either $v_1=1$ or $v_2=1$.    \hfill $\Box$  

\medskip
As rank$\,T = 2$, equation (\ref{e4}) reads $\mathcal{L}_{\mathbf{X}} \phi^\alpha = b \,D^\alpha_{\;\beta} \phi^\beta \,$, whence it follows that $ \mathcal{L}_{\mathbf{X}} \left(\phi^1 \wedge \phi^2\right) =0 \,$ and for the binary wedge products $\Upsilon^\alpha$ we have that  
$\, \mathcal{L}_{\mathbf{X}}\Upsilon^1  = b\, \Upsilon^2 \, $. Then, as $\Upsilon^1 =0$ and $\Upsilon^2 \neq 0$, it follows that $b=0$ and therefore \begin{equation} \label{MC2.19}   
\mathcal{L}_{\mathbf{X}} \phi^\alpha = 0\, , \qquad \alpha = 1,2 
\end{equation}

Now two cases must be separately considered:

\paragraph{Case 2.I.b.1:} If $\,v_{\alpha} \neq 0$, then either $v_\alpha=(1,v)$ or $v_\alpha=(v,1)$. 
Now the integrability conditions for equations (\ref{MC2.19}) imply that $\mathcal{L}_{\mathbf{X}} d\phi^\alpha = 0 $, which leads to $\mathcal{L}_{\mathbf{X}}\Sigma^{\alpha\beta} = \mathcal{L}_{\mathbf{X}} \Upsilon^2 =0$ and therefore
$ \mathbf{X} r = \mathbf{X} s = 0 $.
Substituting this in equation (\ref{MC2.16}) we readily obtain that: 
$$  \mathcal{L}_{\mathbf{X}} \left(\phi^3 \wedge \phi^4 \right)= 0 \qquad {\rm and} \qquad 
  \mathbf{X} v_\alpha \,\phi^\alpha \wedge \phi^3 + v_\alpha \phi^\alpha \wedge  \mathcal{L}_{\mathbf{X}}\phi^3 = 0  $$
which, including that either $v_1=1$ or $v_2=1$, lead to:
\begin{equation} \label{MC2.21}   
\mathcal{L}_{\mathbf{X}} \phi^3 = 0 \qquad {\rm and} \qquad \mathcal{L}_{\mathbf{X}} \phi^4 \wedge \phi^3 = 0
\end{equation}

Equations (\ref{MC2.19}) and (\ref{MC2.21}) can then be unified as
\begin{equation} \label{MC2.22a}
\mathcal{L}_{\mathbf{X}} \phi^a = \delta^a_4 \,f \, \phi^3 \,, \qquad \mbox{for some function } f \,. 
\end{equation}
which is equivalent to: $dX^a = \left( X^e C^a_{ec} + f \delta^a_4 \delta^3_c\right)\,\phi^c$, where as before $\mathbf{X}= X^a \mathbf{Y}_a$ and $C^a_{bc}$ are the commutation coefficients in this base.

This is a first order partial differnetial system on the unknowns $X^a$ but, due to the occurrence of the unknown function $f$, it is not in closed form. However, in some cases the integrability conditions could help to determine $f$. 

The integrability condition for the equation (\ref{MC2.22a}), $a=4$, yields
\begin{equation} \label{MC2.20}
M = f\,\left( C^4_{4\beta} \phi^3\wedge\phi^\beta + d\phi^3\right) + df \wedge \phi^3 \,, \qquad {\rm with} \qquad
M:=-\frac12\,\mathbf{X} C^4_{cb} \phi^c \wedge \phi^b 
\end{equation}
where the fact that $ d\phi^a = - \frac12\, C^a_{bc} \,d\phi^b \wedge d\phi^c \,$ has been included.

Now, if $\phi^3 \wedge d\phi^3 \neq 0$, we can obtain $f=f(X^c)$, which closes the differential system (\ref{MC2.22a}). If it is integrable, then the solution depends on the four real parameters $ X^a_0$, which are submitted to the hyerarchy of constraints that follow from the full integrability conditions of the system (\ref{MC2.22a}), and ${\rm dim}\,\mathcal{C} \leq 4$. 

If, on the contrary, $\phi^3 \wedge d\phi^3 = 0$, 
then it exists $\psi$ such that $d\phi^3= \phi^3\wedge \psi$ and equation (\ref{MC2.20}) implies that $M=\phi^3 \wedge \mu$, for some $\mu$. Moreover, the integrability condition for  equation (\ref{MC2.20}) leads to
$$ f\,\phi^3 \wedge d\left( C^4_{4\beta} \phi^\beta \right) + d M - C^4_{4\beta} M \wedge \phi^\beta = 0   $$
or, separating $f$ in all terms,
\begin{equation}  \label{MC2.24a}
f\,\phi^3 \wedge \left[d\left( C^4_{4\beta} \phi^\beta \right) - \frac12  C^4_{cb|4}\,\phi^c\wedge \phi^b \right] = F_{ecb}( X^a) \,\phi^e\wedge \phi^c\wedge \phi^b  
\end{equation}
which, provided that the right hand side does not vanish, permits to derive $f=f( X^a)$, which closes the partial differential system (\ref{MC2.20}); therefore ${\rm dim}\,\mathcal{C} \leq 4$.

We do not analise the highly non-generic case that neither equation (\ref{MC2.20}) nor equation (\ref{MC2.24a}) can be solved for $f$, which would require furhter study. 

\paragraph{Case 2.I.b.0:} If $\,v_\alpha=0$, then  by (\ref{MC2.16}) we have that 
$$ d\phi^2 =\frac{s}{2}\, \phi^1 \wedge\phi^2 +\phi^3 \wedge\phi^4 \,, \qquad {\rm and }  \qquad d\phi^1 = r \,\phi^1\wedge \phi^2   $$
The exterior derivative of the latter yields $ dr \wedge \phi^1 \wedge \phi^2 - r \phi^1 \wedge \phi^3 \wedge \phi^4 = 0 $, which implies that $r=0$, i. e. $d\phi^1=0$ and locally a function $y$ exists such that $\phi^1= dy$. 
The condition (\ref{MC2.19}) then implies that $\mathbf{X} y = C$, constant, and two cases must be considered depending on whether $s$ does vanish or not:
\begin{description}
\item[2.I.b.0.nd] If \underline{$s\neq 0$}, then $d\phi^2$ is simplectic and we can apply the results in section \ref{S3.1}, case {\bf 1.nd}. Using canonical coordinates, $\phi^2 = p_ i \,dq^i$, $i=1,2$, and $ \mathbf{X} = - \{f,\_\} \,$,  where $\,p_i f^{|i} = f \,$. As a consequence, $f$ is a solution of the partial differential system:
\begin{equation} \label{MC2.25}
 p_i \partial^i f = f\,, \qquad \qquad \{y,f\}  = C\,.
\end{equation}  
In order to study its integrability, consider the minimal integrable submodule $\mathcal{H} \subset \mathscr{X} (\mathcal{M})$  containing $\mathbf{P}=p_j \partial^j$ and $\mathbf{Y}=\{y,\_\}$. It is obvious that $2 \leq {\rm dim}\,\mathcal{H} \leq 4$ and that $df \in \mathcal{H}^\perp$. Therefore,
\begin{itemize}
\item if ${\rm dim}\,\mathcal{H} = 4$, then $df=0$ and there are no $T$-collineations at all, and
\item if ${\rm dim}\,\mathcal{H} < 4$, then $0 <{\rm dim}\,\mathcal{H}^\perp \leq 2$ is the number of arbitrary functions on which $f$ is built of.
\end{itemize}

\item[2.I.b.0.d] If \underline{$s= 0$}, then $d\phi^2\wedge d\phi^2 =0$ but, as $\Upsilon^2\neq 0$, we also have that $d\phi^2\wedge \phi^2 \neq 0$, the results in section \ref{S3.1}, case {\bf 1.d} apply and canonical coordinates can be chosen such that $ \phi^2 = d q^2 + p_1 \,dq^1 \,$, $\,\phi^1 = dp_2$ and
\begin{equation} \label{MC2.28}
 \mathbf{X} = f^{|1} \partial_1 - f_{|1} \partial^1 + (f - p_1 f^{|1})\,\partial_2 + X^2 \partial^2 \,, \qquad f= f(p_1, q^1) 
\end{equation}
with $X_2 = C - f^{|1} \partial_1 y + f_{|1} \partial^1 y - (f - p_1 f^{|1})\,\partial_2 y $, where the condition $i_\mathbf{X} \phi_1 = C\,$ has been included. 
\end{description}

%%%%%%%%%%%%%%%%%%%%%%%%%%%%%%%%%%%%%%%%%%%%%%%%%%%%%%%%%%
\subsection{Type 2.N  \label{S3.2.4}}
This case only occurs when $\sigma= -1$, i. e. $\eta_{\alpha\beta} = {\rm diag}(1,\,-1)$. 
We shall write $\Sigma^{11} = t \,\Upsilon^1 $, $\Sigma^{12} = r \,\Upsilon^1 $ and $\Sigma^{22} = s \,\Upsilon^1 $.

\begin{proposition}  \label{p.MC4}
If $\Upsilon^1 = \Upsilon^2 \neq 0$, then two differential forms, $\phi^3 $ and $\phi^4 $, exist such that 
\begin{equation} \label{MC2.29}
d\phi^1=\frac{t}2\, \phi^1 \wedge \phi^2 + \phi^3 \wedge \phi^4 \,, \qquad   d\phi^2= d\phi^1 + (r-t)\,\phi^1 \wedge \phi^2 + P_{\alpha\beta} \,\phi^\alpha \wedge \phi^{\beta+2}\,.
\end{equation}
with $\det P_{\alpha\beta} = r -(s+t)/2 $ and the differential forms $\phi^{\alpha+2}$ are determined up to the gauge transformation $\phi^{\prime\,\alpha+2} = L^\alpha_{\;\nu} \phi^{\nu+2}$ where $L^\alpha_{\;\nu}$ is a $SL(2)$ valued function.
\end{proposition}

\paragraph{Proof:} Consider $\Theta := d\phi^1 -\frac{t}2\,\phi^1 \wedge \phi^2$. As $\Theta \wedge \Theta = 0$, it is simple and two differential forms, $\phi^3 $ and $\phi^4 $, exist such that $\Theta = \phi^3 \wedge \phi^4$. 
Moreover, as $\Upsilon^1\neq 0$, $\phi^1 \wedge \phi^2 \wedge \phi^3 \wedge \phi^4 \neq 0$. 

Then, as $(d\phi^2 - d\phi^1) \wedge \phi^1 \wedge \phi^2 = 0$, we have that 
$d\phi^2 - d\phi^1 = a\,\phi^1 \wedge \phi^2 + P_{\alpha\beta} \,\phi^\alpha \wedge \phi^{\beta+2}$ and, including the values of $\Sigma^{\mu\nu}$ we derive that $a = r-t$ and $\det(P_{\alpha\beta}) = r -(s+t)/2$.  \hfill $\Box$

\begin{proposition}  \label{p.MC41}
\begin{list}
{(\alph{llista})}{\usecounter{llista}}
\item If $ t+s-2r \neq 0$, then $\phi^3 $ and $\phi^4 $ can be chosen so that 
\begin{equation} \label{MC2.29a}
P_{\alpha\beta} = \sqrt{\frac{|2 r - s - t|}{2}}\, K_{\alpha\beta}\,, \qquad {\rm with} \qquad K_{\alpha\beta} = {\rm diag}(1,\,\kappa)\,, \qquad \kappa={\rm sign}(2 r - s -t)   
\end{equation}
The differential forms $\phi^3 $ and $\phi^4 $ are uniquely determined and $\{\phi^a\}_{a=1\ldots 4}$ is the {\em canonical base} for the tensor $T$. 
\item If $ t+s-2r = 0$, then $\phi^3 $ and $\phi^4 $ can be chosen so that 
$ P_{\alpha\beta} = v_\alpha \delta^1_\beta  $. 
%Besides, the gauge $\phi^{\prime 4} = \phi^4 + m \phi^3$ is left.
\end{list}
\end{proposition}

\paragraph{Proof:}    
\begin{list}
{(\alph{llista})}{\usecounter{llista}}
\item By the gauge  $\phi^{\prime \alpha+2} = L^\alpha_{\;\nu} \phi^{\nu+2}\,,$ where $L^\alpha_{\;\nu} \in SL(2)$, the matrix $P_{\alpha\beta}$ transforms as $P^\prime_{\alpha\beta} = P_{\alpha\nu} L^\nu_{\;\beta}$. 
Therefore, by an appropriate choice of the gauge, we can get $P_{\alpha\beta}$ in the shape (\ref{MC2.29a}).
\item In this case $\det(P_{\alpha\beta})=0$ and therefore $P_{\alpha\beta}= v_\alpha w_\beta$. Then the $SL(2)$ gauge  can be chosen so that $w_\beta L^\beta_{\;\nu} = \delta^1_\nu$. \hfill $\Box$
\end{list}

The value of $\det(P_{\alpha\beta})$ is $T$-frame dependent. Indeed, by a $T$-rotation we have that
\begin{equation} \label{MC2.29e}
\tilde\Sigma^{\alpha\beta} = R^\alpha_{\;\mu}  R^\beta_{\;\nu} \Sigma^{\mu\nu} - d R^\alpha_{\;\mu} \wedge d R^\beta_{\;\nu} \wedge\phi^\mu \wedge \phi^\nu + \left( R^\alpha_{\;\nu}\, d R^\beta_{\;\mu} + R^\beta_{\;\nu} \,dR^\alpha_{\;\mu}\right)   \wedge\phi^\mu \wedge d \phi^\nu \,,
\end{equation}
where $R^\alpha_{\;\nu}$ is a $O(1,1)$ matrix valued function. Using that we obtain
$$ R^\alpha_{\;\nu} = \left(\begin{array}{cc}
                    \cosh \zeta & \sinh \zeta \\  \sinh \zeta & \cosh \zeta 
                    \end{array}  \right) \,, \qquad {\rm and} \qquad 
d R^\alpha_{\;\beta}=  R^\alpha_{\;\nu} D^\nu_{\;\beta}\,d\zeta  $$
whence it easily follows that $d R^\alpha_{\;\mu} \wedge d R^\beta_{\;\nu}=0$. 

Particularly we have that:
$$ \Sigma^{11} + \Sigma^{22} - 2 \Sigma^{12} = e^{-2\zeta}\,\left[ \tilde\Sigma^{11} + \tilde\Sigma^{22} - 2 \tilde\Sigma^{12} - 2 \,d\zeta \wedge (\phi^1-\phi^2) \wedge (d\phi^1-d\phi^2) \right] $$
and, as $\tilde\Upsilon^1 = e^\zeta \,\Upsilon^1 $, this amounts to
$$ e^{3\zeta} \, (\tilde t + \tilde s - 2 \tilde r)\,\Upsilon^1 = ( t + s - 2 r)\,\Upsilon^1 - 2 \,d\zeta \wedge (\phi^1-\phi^2) \wedge d(\phi^1-\phi^2) $$
which finally leads to
\begin{equation} \label{MC2.29c}
e^{3\zeta} \, \left(\frac{\tilde t + \tilde s}{2} -  \tilde r \right) = \left( \frac{t + s}{2} -  r\right) + (P_{24}+P_{14})\,\zeta_3 - 
(P_{23}+P_{13})\,\zeta_4
\end{equation}
where $d\zeta=\zeta_a\,\phi^a$. 

\begin{proposition}  \label{p.MC411}
For type {\bf 2.N} tensors it always exists a $T$-frame such that $ 2  \tilde r - \tilde t - \tilde s = 0$.
\end{proposition}

\paragraph{Proof:} If $2r - t -s \neq 0$ ---case (a) in Proposition \ref{p.MC41}--- and we look for a rotated $T$-frame such that $ 2  \tilde r - \tilde t - \tilde s = 0$, equation (\ref{MC2.29c}) yields 
$$ \kappa \zeta_3 -\zeta_4 + \kappa J  = 0 \,, \qquad {\rm where} \qquad  J:=\sqrt{|r-(s+t)/2}\,,$$
which obviously has many solutions, $\zeta$.  \hfill $\Box$

\begin{corollary}  \label{c.1}
For type {\bf 2.N} tensors it always exists a $T$-frame such that 
\begin{equation} \label{MC2.321a}
d\phi^1=\frac{t}2\, \phi^1 \wedge \phi^2 + \phi^3 \wedge \phi^4 \,, \qquad   d\left(\phi^2-\phi^1\right) = \frac{s-t}{2} \,\phi^1 \wedge \phi^2 + \left[ w\,(\phi^2 - \phi^1) + u \,\phi^2 \right] \wedge \phi^3\,.
\end{equation}
with either {\em (a)} $w=1$ or {\em (b)} $w=0$ and $u=1$ or 0
\end{corollary}

\paragraph{Proof:} It follows easily from propositions \ref{p.MC41} and \ref{p.MC411}; then the gauge freedom in $\phi^3 $ can be used to suitably scale $w$ and $u$. \hfill $\Box$

\begin{proposition}  \label{p.MC42}
If $u=0$ and $s+t=2r$, a $T$- frame can be found such that $\tilde t = \tilde s = \tilde r$.
\end{proposition}

\paragraph{Proof:}
If $u=0$, including equation (\ref{MC2.29c}), we have that $t+s-2r=0$ is invariant by $T$-rotations. Furthermore, using the transformation (\ref{MC2.29e}), we easily arrive at:
$$ \tilde\Sigma^{11} -  \tilde\Sigma^{22} = \Sigma^{11} -  \Sigma^{22} + 2\, d\zeta \wedge\left(\phi^1 \wedge d\phi^2 - \phi^2 \wedge d\phi^1\right) $$
and, as $\tilde\Upsilon^1 = e^\zeta\,\Upsilon^1 $, 
$$ (\tilde t - \tilde s )\,e^\zeta \,\Upsilon^1 = (t-s)\,\Upsilon^1+ 2\, d\zeta \wedge\left(\phi^1 \wedge d\phi^2 - \phi^2 \wedge d\phi^1\right) \,.$$
Therefore, by a suitable choice of $\zeta$ we can make $\tilde t = \tilde s$. \hfill $\Box$

As rank$\,T = 2$, equation (\ref{e4}) reads 
\begin{equation} \label{MC2.30-}
\mathcal{L}_{\mathbf{X}} \phi^\alpha = b \,D^\alpha_{\;\beta} \phi^\beta \,, 
\end{equation}
whence it follows that $ \mathcal{L}_{\mathbf{X}} \left(\phi^1 \wedge \phi^2\right) =0 \,$ and, as $\Upsilon^1 = \Upsilon^2$, we have that $\, \mathcal{L}_{\mathbf{X}}\Upsilon^1  = b\, \Upsilon^1 \, $.

The integrability conditions 
\begin{equation} \label{MC2.30}   
\mathcal{L}_{\mathbf{X}} d\phi^\alpha = D^\alpha_{\;\beta} \,\left(b \, d\phi^\beta + db \wedge \phi^\beta  \right)
\end{equation}
must also be considered.

\begin{proposition}  \label{p.MC511}
The necessary condition for equations (\ref{MC2.30-}) and (\ref{MC2.30}) to have a solution is that 
\begin{description}
\item[(0)] either $u=0$\, and therefore $\phi^2 -\phi^1$ is integrable, and
\begin{equation} \label{MC2.302a}
db = w \mathcal{L}_{\mathbf{X}} \,\phi^3 + \tilde b_\alpha \phi^\alpha  \qquad {\rm or}
\end{equation}
\item[(1)]  $u\neq 0$, in which case
\begin{eqnarray}  \label{MC2.312a}
 & & \mathcal{L}_\mathbf{X}\left( u  \phi^3\right) = - 2 b u \,\phi^3 \\    \label{MC2.312b}
 & & \mathcal{L}_\mathbf{X}\,\left( u^{-1}\,  [\phi^4 - w\,\phi^2]\right) = 3 b\,u^{-1}\left(  \phi^4 - w \,\phi^2\right) - 2b\,\phi^2 + f u \,\phi^3      \\    \label{MC2.312c}
 & &  db = -\left( \mathbf{X}\log u + b[u+2w] \right)\,\phi^3 + \frac12\, \left( \mathbf{X} t - bs\right) \,\phi^1 - \frac12\, \left( \mathbf{X} s - b t\right) \,\phi^2 \,,
\end{eqnarray}
where $f$ is some function. 
\end{description}
\end{proposition}

\paragraph{Proof:} From equation (\ref{MC2.30-}) we have that $\mathcal{L}_{\mathbf{X}} (\phi^2 -\phi^1) = - b (\phi^2 -\phi^1)$, whose  integrability condition implies that: 
\begin{equation} \label{MC2.301}
\mathcal{L}_{\mathbf{X}} \,d(\phi^2 -\phi^1) = - db \wedge (\phi^2 -\phi^1) - b \,d(\phi^2 -\phi^1) 
\end{equation}
Its external product by $\phi^\beta$ yields 
$$ \epsilon^{\beta\alpha} \,\phi^1\wedge\phi^2\wedge\, \left( \left[\delta^2_\alpha \mathbf{X}u\,\phi^3 + b u \,\phi^3\right] + \left[\delta^2_\alpha  u - w V_\alpha \right]\,\mathcal{L}_{\mathbf{X}} \,\phi^3 + V_\alpha \, db  \right)=0  $$
which amounts to:
\begin{equation} \label{MC2.302}
\phi^1\wedge\phi^2\wedge \left[ \mathcal{L}_{\mathbf{X}} \left( u\,\phi^3\right) + 2 b u \,\phi^3 \right]= 0 \qquad {\rm and} \qquad 
\phi^1\wedge\phi^2\wedge \,\left( db -w \mathcal{L}_{\mathbf{X}} \,\phi^3 + b u\,\phi^3 \right)=0  \,, 
\end{equation}
where equation (\ref{MC2.321a}) has been included. 

If $u\neq 0$, the first of these equations means that $\mathcal{L}_{\mathbf{X}} \left( u\,\phi^3\right) = - 2 b u \,\phi^3 + A_\alpha \phi^\alpha$. Besides, the integrability conditions (\ref{MC2.30}) can be further exploited to obtain that
\begin{eqnarray} \label{MC2.31a}  
  & & \mathbf{X}t - bs - 2 b_1 =0  \,, \qquad\qquad \mathbf{X}s - bt + 2 b_2 =0 \\   \label{MC2.31c}
 & & \mathcal{L}_{\mathbf{X}} \left( u\,\phi^3\right) = - 2 b u \,\phi^3 \\   \label{MC2.31d}
 & & \phi^3\wedge \,\left(\mathcal{L}_\mathbf{X} \phi^4 -(3b +\mathbf{X}\log u) \,\phi^4 + \left[w(\mathbf{X} \log u+2b ) +  2b\,u \right]\,\phi^2 + b \, w (\phi^2-\phi^1) \right) = 0 \,.
\end{eqnarray}
Then equations (\ref{MC2.312a}) to  (\ref{MC2.312c}) follow immediately.

If on the contrary $u = 0$, the first of equations (\ref{MC2.302}) is identically satisfied and the other implies equation (\ref{MC2.302a}).
\hfill $\Box$

\paragraph{Subtype 2.N.1:} This corresponds to $u\neq 0$ and equations (\ref{MC2.312a}) to (\ref{MC2.312c}) hold. We take the base $\hat\phi^1 := \phi^1 $, $\hat\phi^2 := \phi^2 $, $\hat\phi^3 := u\,\phi^3 $ and 
$\hat\phi^4 := u^{-1} (\phi^4 - w \phi^2)$, and equations (\ref{MC2.30-}) and (\ref{MC2.312a}) to (\ref{MC2.312c}) can be written as
\begin{equation} \label{MC2.320}
\mathcal{L}_\mathbf{X} \hat\phi^a = \left( b U^a_c + f \delta^a_4 \delta^3_c\right)\,\hat\phi^c \,, \qquad \qquad 
db =  \left( \mathbf{X} B_c + b E_c\right) \,\hat\phi^c 
\end{equation}
where $a,c = 1 \ldots 4$, the only nonvanishing $U^a_c$ are $U^1_2 =U^2_1 = 1$, $U^3_3=U^4_2=-2$ and $U^4_4=3$, and
$$ B_c = \left(\frac12\,t,\,-\frac12\,s,\,\,\frac{w}{u},\,0\right) \,,\qquad  \qquad 
E_c = \left(-\frac12\,s,\,\frac12\,t,\,\,-1 -\frac{2w}{u}\,0\right)   $$

Let $\{\mathbf{Y}_a\}$ be the dual base of $\{\hat\phi^a\}$, with $d\hat\phi^a = \displaystyle{-\frac12\,\hat C^a_{cb} \hat\phi^c\wedge \hat\phi^b}$, and $\mathbf{X}= \hat X^a \mathbf{Y}_a$. The first of equations (\ref{MC2.320}) then reads $d\hat X^a = \left(\hat X^e \hat C^a_{ec} + b U^a_c + f \delta^a_4 \delta^3_c \right) \,\hat\phi^c$. Therefore, equation (\ref{MC2.320}) is a partial differential system on the unknowns $\hat X^a$ and $b$.

Although this PDS is not in closed form, due to the presence of an arbitrary function, the integrability conditions may help to determine $f$. Indeed, after a little algebra, the integrability condition for equation (\ref{MC2.320}) with $a=4$ becomes
\begin{equation}  \label{MC2.321}
M = f\,\left(\hat C^4_{4\beta} \hat\phi^3\wedge\hat\phi^\beta + d\hat\phi^3\right) + df \wedge \hat\phi^3 
\end{equation}
where
$$ M:= -\,\left(\frac12\,\mathbf{X}\hat C^4_{cb} + b\left[\hat C^4_{ab} U^a_c - \frac12\, U^4_a\hat C^a_{cb}  \right] + U^4_b\,\mathbf{X} B_c  \right)\, \hat\phi^c\wedge\hat\phi^b   $$

Now, if $\hat\phi^3 \wedge d\hat\phi^3 \neq 0$, we have that $f \,\hat\phi^3 \wedge d\hat\phi^3 = \hat\phi^3 \wedge M $
and we can obtain $f=f(\hat X^c, b)$, which closes the differential system (\ref{MC2.320}). If it is integrable, then the solution depends on the five real parameters $\hat X^a_0$ and $b_0$. Similarly as in previous cases, these parameters are submitted to the hierarchy of constraints that follow from the full integrability conditions, and ${\rm dim}\,\mathcal{C} \leq 5$. 

If, on the contrary, $\hat\phi^3 \wedge d\hat\phi^3 = 0$, then it exists $\psi$ such that $d\hat\phi^3= \hat\phi^3\wedge \psi$ and equation (\ref{MC2.321}) implies that $M \wedge \hat\phi^3 =0$, that is a $\mu$ exists such that $M=\hat\phi^3 \wedge \mu$. Moreover, the integrability condition for  equation (\ref{MC2.321}) leads to
$$ f\,\hat\phi^3 \wedge d\left(\hat C^4_{4\beta} \phi^\beta \right) + d M - \hat C^4_{4\beta} M \wedge \hat\phi^\beta = 0  $$
or, separating $f$ in all terms,
\begin{equation}  \label{MC2.322}
 f\,\hat\phi^3 \wedge \left[d\left(\hat C^4_{4\beta} \phi^\beta \right) - \left( \frac12 \hat C^4_{cb|4}- 2 B_{c|4}  \delta^2_b + 3 B_{c|4} \delta^4_b\right)\,\hat\phi^c\wedge \hat\phi^b \right] = F_{ecb}(\hat X, b) \,\hat\phi^e\wedge \hat\phi^c\wedge \hat\phi^b  
\end{equation}
which, provided that the right hand side does not vanish, permits to derive $f=f(\hat X^c,b)$ and then ${\rm dim}\,\mathcal{C} \leq 5$.

Similarly as in case {\bf 2.I.b.1} above, we do not consider the residual nongeneric subcase that neither equation (\ref{MC2.321}) nor equation (\ref{MC2.322}) can be solved for $f$. 

\paragraph{Subtype 2.N.0:} This corresponds to $u=0$ and $w= 0$ or 1 and, including Proposition \ref{p.MC42}, equations (\ref{MC2.321a}) read
\begin{equation} \label{MC2.304}   
d\phi^1=\frac{t}2\, \phi^1 \wedge \phi^2 + \phi^3 \wedge \phi^4 \,, \qquad   d\left(\phi^2-\phi^1\right) = w\,(\phi^2 -\phi^1)\wedge \phi^3 \,,
\end{equation}
Therefore $\phi^2 -\phi^1$ is integrable and, as $\Sigma^{11} \neq 0$, $d\phi^1$ is symplectic, canonical charts $(p_i,\,q^j)$, with $i,j=1,2$, and two functions $u$ and $y$ exist such that 
\begin{equation} \label{MC2.305}   
\phi^1 = p_i\,dq^i \,, \qquad \qquad \phi^2 -\phi^1 = y\,du\,, \qquad \qquad y > 0
\end{equation}

Equation (\ref{MC2.30-}) then leads to $\mathcal{L}_\mathbf{X} (\phi^2 -\phi^1) = - b\, (\phi^2 -\phi^1) $ which, written in terms of $u$ and $y$, implies that a one variable function $B(u)$ exists such that
\begin{equation} \label{MC2.306}   
\mathbf{X} u = B(u) \,, \qquad \qquad - b y = \mathbf{X} y + y B^\prime(u)
\end{equation}
where $B^\prime$ means the derivative. 

From equation (\ref{MC2.30-}) we also have that $\mathcal{L}_\mathbf{X} \phi^1 = b\phi^2$, which in terms of canonical coordinates reads
\begin{equation} \label{MC2.307}   
\left(X_i - b p_i\right)\,dq^i + p_j \,dX^j - y b\,du = 0
\end{equation}
where we have written $\mathbf{X}= X^j\partial_j + X_i \partial^i$  (as in section \ref{S3.1}). Their components are
\begin{eqnarray} \label{MC2.308a}
 & & X_i + v_i \,\left( y B^\prime + \mathbf{X}y \right) + p_j\,\partial_i X^j = 0 \,, \qquad {\rm with} \qquad 
 v_i := \frac1{y}\,p_i + u_{|i} \\  \label{MC2.308b}
 & & p_j \,\partial^i X^j + \left( y B^\prime + \mathbf{X}y \right)\, u^{|i} = 0
\end{eqnarray}
where $u_{|i}:=\partial_i u$ and  $u^{|i}:=\partial^i u$.

If we now  write the components $X^j$ as 
\begin{equation} \label{MC2.308c} 
X^j = \frac1{z^2}\,\left(\xi p^j + \eta\,r^j\right) \,, \qquad {\rm with} \quad p^j:=p_j\,, \quad  r^j:=r_j =(p_2,-p_1)\quad {\rm and} \quad z^2 :=p_i p^i \,,
\end{equation}
equation (\ref{MC2.308a}) becomes
\begin{equation} \label{MC2.308d}   
X_j\,\left(\delta_i^j + v_i y^{|j}\right) + \partial_i\xi + v_i\,\left(y B^\prime + X^j y_{|j}\right)= 0
\end{equation}

\paragraph{Case 2.N.01:} If $1+ v_l y^{|l} \equiv y(1 + y^{|l} u_{|l})+ p_l y^{|l} \neq 0$, we can derive:
\begin{equation} \label{MC2.308}   
X_i = -\frac{y B^\prime + X^j y_{|j}}{1 + v_k y^{|k}}\, v_i - p_j\,\partial_l X^j\,\left( \delta^l_i - \frac{ v_i y^{|l}}{1 + v_k y^{|k}}\right)\,,
\end{equation}
Now equations (\ref{MC2.308b}) together with the first of equations (\ref{MC2.306}) are to be taken as a partial differential system on the two unknowns $X^j$. 
Using equation (\ref{MC2.308}) and after a little algebra, this PDS can be written as
\begin{eqnarray}  \label{MC2.310a}
& & \partial^i \xi - \frac{u^{|i}}{1 + v_k y^{|k}}\,y^{|l} \partial_l\xi - X^i + u^{|i}\,\frac{yB^\prime +
 X^j y_{|j}}{1 + v_k y^{|k}} = 0\\  \label{MC2.310b}
& &  \left(u^{|l}-  \frac{u^{|j} v_j}{1 + v_k y^{|k}}\,y^{|l} \right)\,\partial_l\xi - B - 
\frac{y B^\prime  v_l u^{|l}}{1 + v_k y^{|k}} + X^j\,\left(u_{|j} - \frac{v_l u[^{|l}}{1 + v_k y^{|k}}\,y_{|j}\right) = 0
\end{eqnarray}

Including now the decomposition (\ref{MC2.308c}), $\eta$ can be derived from one of these equations whenever 
\begin{equation} \label{MC2.308g}   
\left(p_j u^{|j}\right)\,\left(r^l y_{|l}\right) \neq 0 \qquad {\rm or} \qquad 
\left(1 + v_k y^{|k}\right) \,(r^j u_{|j}) \neq \left(v_k u^{|k}\right) \,(r^j y_{|j}) \,,
\end{equation}
the remaining two equations then yielding a PDS to be fulfilled by the unknown $\xi$.

On the contrary, if none of the above inequalities hold, $\eta$ is arbitrary, does not occur in the PDS and we are left with three equations on the unknown $\xi$. 

In any case, the PDS looks like:
$$ \mathbf{H}_\alpha \xi = m_\alpha \xi + n_\alpha \,, $$
with  $\alpha$ running  either from 1 to 2 (resp.,  1 to 3). Using the commutation relations we then find the minimal integrable modulus $\mathcal{H}$ containing the fields $\mathbf{H}_\alpha$. 

The solution $\xi$ then depends on an arbitrary function of $ 4- {\rm dim\,}\mathcal{H}$ variables. 
The component $\eta $ is either determined or arbitrary, depending on whether the inequalities (\ref{MC2.308g}) do hold or do not, and the components $X_j$ can be derived from  (\ref{MC2.308}).

\paragraph{Case 2.N.00:} In case that $y(1 + y^{|l} u_{|l})+ p_l y^{|l} = 0$, equation (\ref{MC2.308d}) implies the constraint 
\begin{equation} \label{MC2.308e}   
y^{|l}\partial_l\xi= y B^\prime + X^j y_{|j}
\end{equation}
and its general solution is 
\begin{equation} \label{MC2.308h}   
X_i = -\partial_i \xi + \zeta v_{|i}
\end{equation}
the component $\zeta$ being arbitrary. Including these, equations (\ref{MC2.308b}) and (\ref{MC2.306}) become
\begin{eqnarray} \label{MC2.309a}   
 & & \partial^i \xi - X^i - \zeta u^{|i} = 0  \\  \label{MC2.309b}   
 & & u^{|l}\partial_l \xi = X^l u_{|l} + \zeta v_l u^{|l} - B
\end{eqnarray}

Now, if $u^{|i} \neq 0$, the first of these equations permits to obtain 
\begin{equation} \label{MC2.309c}   
 \zeta = \left(\delta_{ij} u^{|i} u^{|j} \right)^{-1}\,\delta_{kl} u^{|k}\,\left(\partial^l\xi - X^l \right) \,, 
\end{equation}
which substituted in equations (\ref{MC2.308e}), (\ref{MC2.309a}) and (\ref{MC2.309b}) yields a  PDS to be fulfilled by  $\xi$. The discussion about its solution is then similar to that in case {\bf 2.N.01} above. 

If, on the contrary $u^{|i} = 0$, after a little algebra equations (\ref{MC2.308e}), (\ref{MC2.309a}) and (\ref{MC2.309b}) yield:
$\eta = r_{l} \partial^l\xi $ and the PDS:
\begin{equation} \label{MC2.311a} 
p_i \partial^i \xi = \xi \, ,\qquad\qquad  u_{|i} \partial^i \xi = B \, ,\qquad\qquad y^{|l} \partial_l \xi - y_{|l} \partial^l \xi = y B^\prime   
\end{equation}
The discussion about the existence of a solution is then similar to that in  case {\bf 2.N.01} above. 

%%%%%%%%%%%%%%%%%%%%%%%%%%%%%%%%%%%%%%%%%%%%%%%%%%%%%
\subsection{Type 2.H \label{S3.2.5} }
In this case, $\Upsilon^\alpha=0$ and $T$ is holonomous. Therefore coordinates $x^a$, $a=1\ldots 4$, exist such that 
$T= T_{\alpha\beta}(x^a)\, dx^\alpha\otimes dx^\beta$, with $\det T_{\alpha\beta} \neq 0$, and three cases must be separately considered depending on $m:= {\rm rank}\,\{dx^1,\, dx^2,\, dT_{\alpha\beta}\}$, which ranges from 2 to 4:
\begin{description}
\item[Case 2.H.0] If $m=2$, then $\partial_AT_{\alpha\beta}= 0$, $\, A=3,\,4$.
\item[Case 2.H.1] If $m=3$, the coordinates can be chosen so that 
$$ T_{11} = x^3\,, \qquad T_{12}=u\,, \qquad T_{22}= v \qquad {\rm with} \qquad \partial_4 u =\partial_4v =0 $$
\item[Case 2.H.2] If $m=4$, the coordinates can be chosen so that $\, T_{11} = x^3\,, \quad T_{12}=u\,, \quad T_{22}= x^4 \,$. 
\end{description} 

Type {\bf 2.H} tensors will be dealt in much the same way as rank 3 tensors. We first write the collineation field as
$$\mathbf{X} = \mathbf{Z} + f^A\,\mathbf{N}_A \,\qquad \mbox{where $f^A$ are two functions}\,, A=3,4 \,, \qquad  
\mathbf{N}_A = \partial_A   $$
and $\mathbf{Z} = Z^\alpha \partial_\alpha \,$ is tangent to the submanifolds $x^B=\,$constant, $B=3,4$. 

It is obvious that $T( \mathbf{N}_A,\,\_)=0$, which implies that $\mathcal{L}_{f^A\mathbf{N}_A} T = f^A\,\mathcal{L}_{\mathbf{N}_A} T $ and therefore, equation (\ref{e1}) amounts to  
\begin{equation}\label{MC3.H2}
[\mathbf{N}_A,\mathbf{Z}] = 0 \qquad {\rm and } \qquad \mathcal{L}_{\mathbf{Z}} T + 2 f^A K_A = 0  
\end{equation}
where $K_A := \frac12\,\partial_A T_{\alpha\beta} dx^\alpha \otimes dx^\beta $ or, in components,
\begin{equation}\label{MC3.H3}
 \partial_A Z^\alpha = 0 \qquad {\rm and } \qquad \nabla_{(\alpha} Z_{\beta)} +  f^A K_{A|\alpha\beta}= 0
\end{equation}
where $\nabla$ is the Levi-Civita connection for the non-degenerate metric $T_{\alpha\beta}$ on the hypersurfaces $x^B={\rm constant}$.
The second of these equations looks like a non-homogeneous Killing equation (parametrized with $x^B$\,) and  the question is: does it admit solutions $Z^\alpha$ that do not depend on $x^B$ for some appropriate $f^A$?  

In case {\bf 2.H.0}, $K_{A|\alpha\beta}=0$, $\,A=3,4$, the answer is obvious because coordiantes $x^3$ and $x^4$ are mere parameters and equation (\ref{MC3.H3}) reduces to a Killing equation in 2 dimensions. The collineation field is then $\mathbf{X}= \mathbf{Z} + f^A\,\mathbf{N}_A$, where $f^A$ are arbitrary and $\mathbf{Z}$ is a Killing vector for the non-degenerate rank 2 metric $T$ in each submanifold $x^B = {\rm constant}\,$. 

In case {\bf 2.H.1}, $K_{3|\alpha\beta}\neq 0$ and $K_{4|\alpha\beta} = 0$. Then equations (\ref{MC3.H3}) do not involve the function $f^4$, which is arbitrary. The coordinate $x^4$ is only a parameter and the problem has reduced to finding the collineation fields of a rank two tensor on each submanifold $x^4 = {\rm constant}$, which is similar to the problem treated in section \ref{S2}.

The generic case is {\bf 2.H.2}, i. e. $K_{A|\alpha\beta} \neq 0$, $\,A=3,4$. Similarly as in section \ref{S2}, equations (\ref{MC3.H3}) imply that:
\begin{equation}\label{MC3.H4}
\nabla_{\alpha} Z_{\beta} = \Omega_{\alpha\beta} -  f^A K_{A|\alpha\beta} \qquad {\rm with} \qquad  \Omega_{\alpha\beta}+ \Omega_{\beta\alpha}=0
\end{equation}
And the successive integrability conditions that follow from the commutation relations for $\nabla_\alpha$ and $\nabla_\mu$ imply a hierarchy of new equations on $\Omega_{\alpha\beta}$ and $f$, namely
\begin{eqnarray}\label{MC3.H7}
\nabla_\mu \Omega_{\kappa\lambda} & = & R\,Z_{[\lambda} T_{\kappa]\mu} + 2 \nabla_{[\lambda} \left( f^A K_{A|\kappa]\mu}\right)  \\  \label{MC3.H8}
\mathbf{Z} R & = & R \,f^A K_{A|\alpha\beta} T^{\alpha\beta} + 2 \nabla^\alpha \nabla_\alpha \left(f^A K_{A|\nu\beta} T^{\nu\beta}\right) - 2 \nabla^\alpha \nabla^\nu \left(f^A K_{A|\alpha\nu}\right) 
\end{eqnarray}
and so on, where we have included that, as the dimension is two, $R_{\alpha\mu\kappa\lambda} = R \, T_{\lambda[\alpha} T_{\mu]\kappa}\,$.

As for the commutation of the derivatives $\partial_A$ and $\nabla_\alpha$ applied to $Z_\alpha$, we readily obtain that:
\begin{eqnarray}   \label{MC3.H10a}
 & & \partial_A \Omega_{\alpha\beta} = 2 \Omega_{\lambda[\beta} K^{\hspace*{1em}\lambda}_{A|\;\;\alpha]} + 2 Z^\nu \nabla_{[\alpha} K_{A|\beta]\nu} - 2 f^B K_{A|\nu[\alpha} K_{B|\;\;\beta]}^{\hspace*{1em}\nu} \\   \label{MC3.H10b}
 & & \partial_A f^B K_{B|\alpha\beta} + f^B \,\left(\partial_A K_{B|\alpha\beta} - 2 K_{A|\nu(\alpha} K_{B|\;\;\beta)}^{\hspace*{1em}\nu}  \right) + Z^\nu \nabla_\nu K_{A|\alpha\beta} + 2 K_{A|\nu(\alpha} \Omega_{\beta)}^{\;\;\nu} = 0 
\end{eqnarray}

In the case {\bf 2.H.2} we have that
$$ T_{\alpha\beta} = \left( \begin{array}{cc} 
                      x^3 & u \\ u & x^4
                      \end{array}  \right)\,, \qquad 
   T^{\alpha\beta} = \frac1{\Delta}\,\left( \begin{array}{cc} 
                      x^4 & -u \\ -u & x^3
                      \end{array}  \right)\,, \qquad {\rm with} \qquad  \Delta := x^3 x^4 - u^2  $$
and 
$$ K_{A|\alpha\beta} = \frac12\,u_{|A} E_{\alpha\beta} + \delta^{A-2}_\alpha \delta^{B-2}_\beta \,,\qquad {\rm with} \qquad 
E_{\alpha\beta} =\left( \begin{array}{cc} 
                      0 & 1 \\ 1 & 0
                      \end{array}  \right)\,.$$ 
Then equation (\ref{MC3.H10b}) yields
\begin{equation}  \label{MC3.H10c}
2\,\partial_A f^B u_{|B} + f^B N_{AB} = L_A(Z,\Omega) \,, \qquad 
\partial_A f^B + f^C N^B_{AC} = L^B_A(Z,\Omega) 
\end{equation}
with
\begin{eqnarray*}
N_{AB} &:= & 2u_{|AB}+\frac1{\Delta}\,\left(2 u u_{|A} u_{|B} - u_{|B} x^{A^\prime} - u_{|A} x^{B^\prime}    \right)   
\,, \\
 N^B_{AC} &:= &- \frac1{\Delta}\,\left( u_{|A} u_{|C} x^B - u \,u_{|A} \delta^B_C - u \,u_{|C} \delta^B_A 
+ \delta^B_A \delta^B_C x^{B^\prime}   \right)  \,,
\end{eqnarray*}   
$L_A$ and $L^B_A$ are linear functions of $Z^\alpha$ and $\Omega_{\alpha\beta}$, a ``stroke'' means ``partial derivative'' and $A^\prime \neq A$.

Equations (\ref{MC3.H7}) and (\ref{MC3.H10a}) give all derivatives of $\Omega_{\alpha\beta}$ in terms of $Z^\alpha$, $\Omega_{\mu\nu}$, $f^B$ and $\partial_\nu f^B$,and equations (\ref{MC3.H10c}) 
can be taken as a linear system of six equations for the six unknowns $f^B$ and $\partial_A f^B$. If the matrix of the system has rank six, this is a Cramer's system and we can derive 
$$ f^B = F^B(Z^\nu,\Omega_{\alpha\beta}) \,, \qquad  \partial_A f^B = F^B_{\;A}(Z^\nu,\Omega_{\alpha\beta})  $$
Substituting the above relations in equations (\ref{MC3.H3}), (\ref{MC3.H7}) and (\ref{MC3.H10a}), we obtain a closed partial differential system on $Z^\nu$ and $\Omega_{\alpha\beta}$ whose solutions are parametrized by three real parameters, namely the values of $Z^\nu$ and the skewsymmetric $2\times 2$ matrix $\Omega_{\alpha\beta}$ at one point. 

Of course, some constraints will follow from the fact that $F^B_{\;A} = \partial_A F^B $. These, together with the hierarchy of integrability conditions, will result in a homogeneous linear system of conditions on the parameters 
$Z^\nu(0)$ and $\Omega_{\alpha\beta}(0)$.  
Therefore $\mathcal{C}_T$ is  a Lie algebra and ${\rm dim}\, \mathcal{C}_T \leq 3$. 

If the linear system (\ref{MC3.H10c}) is not Cramer's, we can at least derive $\partial_A f^B = \tilde{L}^B_A (Z,\Omega, f^C)$. Now, (\ref{MC3.H7}) and (\ref{MC3.H10a}) give all derivatives of $\Omega_{\alpha\beta}$ and therefore some integrability conditions will follow, namely
\begin{eqnarray}  
\lefteqn{ \epsilon^{\beta\alpha}\,\left( \partial_\beta f^A\,\left[ - N^C_{BA} K_{C|\alpha\mu} + \partial_B K_{A|\alpha\mu} +  K_{B|\nu\alpha} K_{A|\;\;\mu}^{\hspace*{1em}\nu}\right] \right.}  \nonumber \\   \label{MC3.H10d}
& & \hspace*{4em}\left. + \partial_\mu f^A\, K_{B|\nu\beta}    K_{A|\;\;\alpha}^{\hspace*{1em}\nu} - \partial_\nu f^A\, K_{B|\;\;\beta}^{\hspace*{1em}\nu} K_{A|\alpha\mu}\right) = W_{B\mu} 
\end{eqnarray}
where $W_{B\mu}$ is a linear function of $Z^\nu$, $\Omega_{\alpha\beta}$ and $f^B$. This is to be seen as a linear system of four equations on the four unknowns $\partial_\nu f^B$ and, in case that the rank is four, we can derive $\partial_\nu f^B = F^B_\beta /Z,\Omega,f^A)$ which, together with $\partial_A f^B = \tilde{L}^B_A (Z,\Omega, f^C)$ and equations (\ref{MC3.H3}), (\ref{MC3.H7}) and (\ref{MC3.H10a}) yield a closed partial differential system on the unknowns $Z^\nu$, $\Omega_{\alpha\beta}$ and $f^B$. The general solution is a vector space whose dimension is at most five. Therefore $\mathcal{C}_T$ is a Lie algebra and  ${\rm dim}\,\mathcal{C}_T \leq 5$. 

In the case {\bf 2.H.1} we shall do similarly and, as $K_{4|\alpha\beta}=0$, the component $f^4$ does not occur in any equation and is arbitrary. Therefore $\mathcal{C}_T = \mathcal{C}_T^{(0)} + {\rm span} \{\partial_4\}$, where the vector fields in $\mathcal{C}_T^{(0)} $ are characterized by $\mathbf{X} x^4 =0$.

Equation (\ref{MC3.H10b}) also impies that $\partial_4 f^3 =0$ and that: 
\begin{equation}  \label{MC3.H10e}
  \partial_3 f^3 K_{3|\;\;\beta}^{\hspace*{.7em}\alpha} + f^3 \,\partial_3 K_{3|\;\;\beta}^{\hspace*{.7em}\alpha} + Z^\nu \nabla_\nu K_{3|\;\;\beta}^{\hspace*{.7em}\alpha} - K_{3|\;\;\nu}^{\hspace*{.7em}\alpha} \Omega^\nu_{\;\beta)} + \Omega^\alpha_{\;\nu}K_{3|\;\;\beta}^{\hspace*{.7em}\nu}  = 0
\end{equation}
If  $K_{3|\;\;\beta}^{\hspace*{.7em}\alpha} $ and $\partial_3 K_{3|\;\;\beta}^{\hspace*{.7em}\alpha} $ are independent, then we can derive 
$$ f^3 = F(Z^\nu,\Omega_{\alpha\beta}) \,, \qquad  \partial_3 f^3 = F_{\;3}(Z^\nu,\Omega_{\alpha\beta})  $$
and close the partial differential system  (\ref{MC3.H3}), (\ref{MC3.H7}) and (\ref{MC3.H10a}). Its solutions depending on the three real parameters $Z^\nu(0)$ and $\Omega_{\alpha\beta}(0)$, which are further constrained by the hierarchy of integrability conditions, the space $\mathcal{C}_T = \mathcal{C}_T^{(0)}$ is a Lie algebra whose dimension is at most three. 

If, on the conbtrary,  $K_{3|\;\;\beta}^{\hspace*{.7em}\alpha} $ and $\partial_3 K_{3|\;\;\beta}^{\hspace*{.7em}\alpha} $ are not independent, then as $K_{3|\;\;\beta}^{\hspace*{.7em}\alpha} \neq 0$, we can at least derive $\partial_3 f^3 = F_3(Z,\Omega,f^3)$. Now the integrability conditions that follow from equations (\ref{MC3.H7}) and (\ref{MC3.H10a}) yield a linear system of two equations on $\partial_\nu f^3$. Generically this is a Cramer's system and can be solved for to derive $\partial_\nu f^3 = F_\nu(Z,\Omega,f)$ and, together with equations  (\ref{MC3.H3}), (\ref{MC3.H7}) and (\ref{MC3.H10a}), finally close a partial differential system on $Z^\nu$, $\Omega_{\alpha\beta}$ and $f^3$. The general solution depends on four real parameters, namely $Z^\nu(0)$, $\Omega_{\alpha\beta}(0)$ and $f^3(0)$, which are further constrained by the hierarchy of integrability conditions. Therfore  the space $\mathcal{C}_T^{(0)}$ is a Lie algebra whose dimension is at most four. 

%%%%%%%%%%%%%%%%%%%%%%%%%%%%%%%%%%%%%%%%%%%%%%%%%
\section{Summary}
We finally present an outline of the classification of covariant second order tensors, according to their respective classes of collineation fields, $ \mathcal{C}_T$, and summarize what has been proved along previous sections. As a rule, it seems that for generic cases $\mathcal{C}_T$ is a finite dimensional Lie algebra, whereas in nongeneric cases, i. e. some equalities do hold, $\mathcal{C}_T$ is not a Lie algebra and has an infinite number of dimensions. 

\paragraph{Rank 4 tensors:} $T$ can be viewed as a non-degenerate metric on $\mathcal{M}$, $\mathcal{C}_T$ is the corresponding Killing algebra and ${\rm dim}\,\mathcal{C}_T \leq 10$.

\paragraph{Rank 3 tensors:} Local charts exist such that
$ T = T_{\alpha\beta}(y^a)\, dy^\alpha \otimes dy^\beta \,$, $a= 1\ldots 4\,$, $\alpha,\beta = 1 \ldots 3 $. Write then $\mathbf{X}= \mathbf{Z} + f\,\mathbf{N}$, where $T(\mathbf{N},_)=0$, and consider $K_{\alpha\beta} := \mathcal{L}_\mathbf{N}T_{\alpha\beta}$.
\begin{itemize}
\item If $K_{\alpha\beta} =0$, then the collineation fields  are: $\mathbf{X} = \mathbf{Z} + X^4 \partial_4$,
with $X^4$ arbitrary and $\mathbf{Z} y^4 =0$. $\mathcal{C}_T$ is not a lie algebra but the subclass $\mathcal{C}^0_T = \{\mathbf{X}\in \mathcal{C}_T\,|\; \mathbf{X}y^4 =0 \}$ is a Lie algebra that has at most six dimensions.
\item If $\partial_4 K_{\alpha\beta}$ is not proportional to $ K_{\alpha\beta}$, then $\mathcal{C}_T$ is a Lie algebra and ${\rm dim} \mathcal{C}_T \leq 6$.
\item If $\partial_4 K_{\alpha\beta} \propto K_{\alpha\beta}$, but $K^{\alpha\beta} K_{\beta\mu} \neq 0$ and $K_{\alpha\beta} $  is not proportional to $T_{\alpha\beta}$, then $\mathcal{C}_T$ is a Lie algebra and ${\rm dim} \mathcal{C}_T \leq 7$.
\end{itemize}
Our analysis of two residual, degenerate cases, has been left incomplete and they probably involve arbitrary functions, i.e. $\mathcal{C}_T$ is infinite dimensional.
These cases correspond to $\partial_4 K_{\alpha\beta} \propto K_{\alpha\beta}$ and, either $K^{\alpha\beta} K_{\beta\mu} = 0$ or $K_{\alpha\beta} \propto T_{\alpha\beta}$.

\paragraph{Rank 1 tensors:} We write $T= \phi \otimes \phi$ and distinguish several cases:
\begin{description}
\item[Type 1.nd]  $d\phi$ is simplectic and, in canonical coordinates $(q^i, p_j)$, the collineation fields are 
$$ \mathbf{X} = f^{|i} \partial_i - fd_{|j} \partial^j $$
where $f(q^i,p_j)$ is homogeneous and of first degree on the ``momenta'' $p_j$. 
\item[Type 1.d] Characterised by $d\phi\wedge d\phi = 0$ and $d\phi\wedge \phi \neq 0$. Then coordinates $(q^i,p_j)$ exist such that 
$$ \mathbf{X} = f^{|1} \partial_1 - f_{|1} \partial^1 + (f - p_1 f^{|1} )\,\partial_2 + X_2 \partial^2 $$
where $f(p_1,q^1)$ and $X_2(p_i,q^j)$ are arbitrary.
\item[Type 1.dh] Characterised by $d\phi\wedge d\phi = d\phi\wedge \phi = 0$; coordinates $(q^i,p_j)$ exist such that 
$$ \mathbf{X} = F\,\partial_1 - p_1 F^\prime \partial^1 + X^2\partial_2 + X_2 \partial^2   \,, $$
where $F(q^1)$, $X^2(p_i,q^j)$ and $X_2(p_i,q^j)$ are arbitrary functions.
\item[Type 1.d0] In this case $d\phi=0$ and coordinates $x^a$, $a=1 \ldots 4$ exist such that
$$ \mathbf{X} = C \partial_1 + \sum^4_{i=2} X^i \partial_i $$
where $C$ is a constant and the three functions $X^i(q^j,p_l)$ are arbitrary. 
\end{description}

\paragraph{Rank 2 tensors:} We take the canonical expression $T= \eta_{\mu\nu} \phi^\mu \otimes \phi^\nu$, and classify $T$ on the basis of the volume forms $\Sigma^{\mu \nu} = d\phi^\mu \otimes d\phi^\nu$ and $\Upsilon^\mu = d\phi^\mu \otimes \phi^1  \otimes \phi^2$.
\begin{description}
\item[Type 2.I.a] There is a $T$-frame in which $\Sigma^{11} \neq 0$, $\Upsilon^1 = 0$ and $\Upsilon^2 \neq 0$ and a canonical base $\{\phi^a\}_{a=1\ldots 4}$ exists such that $\mathcal{L}_\mathbf{X} \phi^a =  0$.
In this case ${\rm dim}\, \mathcal{C}_T \leq 4$.

\item[Type 2.I.b] There is a $T$-frame in which $\Sigma^{11} = \Upsilon^1 = 0$ and $\Upsilon^2 \neq 0$. Then, by Proposition \ref{p.MC3}, it exists a base, $\phi^\alpha$, $ \alpha = 1 \ldots 4$, in which the differential forms $d\phi^\alpha$ have the canonical expression  (\ref{MC2.16}). Two cases arise according to the values of $v_\alpha$:
   \begin{description}
   \item[Case 2.I.b.1] If $v_\alpha \neq 0$, then it exists a base in which $\mathcal{L}_\mathbf{X} \phi^a = f \delta^a_4\,\phi^3$, for some $f$.
   
   If $\phi^3\wedge d\phi^3 \neq 0$ or $\phi^3 \wedge d\left(C^4_{4b} \phi^b -\frac12\,C^4_{cb|4}\phi^c \wedge \phi^b \right) \neq 0$, then it results that $\mathcal{C}_T$ is a Lie algebra and ${\rm dim}\,\mathcal{C}_T \leq 4$. Otherwise $\mathbf{X}$ might contain arbitrary functions and therefore $\mathcal{C}_T$ is not a Lie algebra has an infinite number of dimensions.
   \item[Case 2.I.b.0.nd]  If $v_\alpha = 0$ and $\Sigma^{22} \neq 0$, then $\phi^1 = dg$ and $d\phi^2$ is simplectic; in canonical coordinates the collineation field is $ \;\mathbf{X} = f^{|i} \partial_i - f_{|i} \partial^i\,$, where $f(q^i,p_j)$ is a solution of the Pfaff system $\mathcal{H}^\perp$ defined in (\ref{MC2.25}). The collineation field $\mathbf{X}$ might depend on an arbitrary function of two variables at most.
   \item[Case 2.I.b.0.d]  If $v_\alpha = \Sigma^{22} = 0$, then canonical coordinates exist in which $\mathbf{X}$ is given by equation (\ref{MC2.28}) and it contains an arbitrary function of two variables.
   \end{description}

\item[Type 2.N] This case, $\Upsilon^1 = \Upsilon^2 \neq 0$, only occurs if the tensor $T$ has no definite sign. According to Proposition \ref{p.MC4}, a canonical base $\{\phi^a\}_{a=1 \ldots 4}$ exists such that the exterior derivatives $d\phi^\alpha$ are given by equation (\ref{MC2.321a}) with either $u \neq 0$ or $u=0$ and $\Sigma^{11}=\Sigma^{22}$. 

\begin{description}
\item[Subtype 2.N.1] If $u\neq 0$, $\mathbf{X}$ is the solution of the partial differential system (\ref{MC2.320}),
which involves an arbitrary $f$. Thus the system is not closed and, provided that either $\hat\phi^3 \wedge d\hat\phi^3 \neq 0$ or that equation (\ref{MC2.322}) can be solved for $f$, the class $\mathcal{C}_T$ is a Lie algebra and ${\rm dim}\,\mathcal{C}_T\leq 5$. 

In the residual nongeneric case that $\hat\phi^3 \wedge d\hat\phi^3 = 0$ and that equation (\ref{MC2.322}) cannot be solved for $f$, the partial differential system might not close and $\mathbf{X}$ might contain arbitrary functions. Therefore $\mathcal{C}_T$ might not a Lie algebra and have an infinite number of dimensions.

\item[Subtype 2.N.0] This corresponds to, $u = 0$ and $w=0$ or 1, then canonical coordinates exist such that 
$$ \phi^1 = p_i \,dq^i \,, \qquad \qquad \phi^2 -\phi^1 = y\,du \,, \qquad y>0 \,. $$
An arbitrary one variable function $B(u)$ appears. 
   \begin{description}
   \item[Case 2.N.01] Characterized by $y ( 1 + u_{|l} y^{|l}) + p_l y^{|l} \neq 0$. The components $X_j$ are determined by equation (\ref{MC2.308}). 
   As for the components $X^j$, if one of the inequalities (\ref{MC2.308g}) 
   holds, then $\eta:=p_1 X^2 -p_2 X^1$ is determined in terms of $\xi:= p_j X^j$, which is a solution of a 2-equations linear partial differential system and, provided that it is integrable, $\xi$ is determined up to the addition of an arbitrary function of two variables at most.
   If no inequality (\ref{MC2.308g})
   holds, then $\eta$ is arbitrary and $\xi$ is a solution of a 3-equations linear partial differential system and, provided that it is integrable, $\xi$ is determined up to the addition of an arbitrary function of one variable at most.
   
   \item[Case 2.N.00] This is characterized by $y ( 1 + u_{|l} y^{|l}) + p_l y^{|l} = 0$ and if $u^{|l} \neq 0$, the components $X_j$ are given by (\ref{MC2.308h}) and (\ref{MC2.309c}).
   The component $ \eta$ is determined by (89) and the component $\xi$ is a solution of equations (\ref{MC2.308e}) and (\ref{MC2.309b}), a linear partial differential system and, provided that it is integrable, $\xi$ is determined up to the addition of an arbitrary function of two variables at most.
   
   If on the contrary $u^{|l} \neq 0$, then $X_j$ are given by (\ref{MC2.308h}) and include an arbitrary function, besides $\eta = p_1\partial^2\xi -p_2 \partial^1\xi$ and $\xi$ is a solution of the linear partial differential system (\ref{MC2.311a}) and, provided that it is integrable, it is determined up to the addition of an arbitrary function of one variable at most.
   \end{description}
Therefore the class of collineation fields is infinite dimensional and is not a Lie algebra.
\end{description}

%%%%%%%%%%%%%%%%%%%%%%%%%%%%%%%%%%%%%%%%%%%%%

\item[Type 2.H] If $\Upsilon^1 = \Upsilon^2 =0$, local charts exist such that $T= T_{\alpha\beta} dx^\alpha\otimes dx^\beta$, $a= 1\ldots 4$, $\alpha, \beta =1,2$. Writing then $\mathbf{X} = \mathbf{Z} + f^A \partial_A$, with $\mathbf{Z}= Z^\nu \partial_\nu$, and  $K_{A|\alpha\beta}:= \frac12\,\partial_AT_{\alpha\beta}$, $A=3,4$,

\begin{description}
\item[Case 2.H.0] If $K_{A|\alpha\beta}=0$, then $\mathcal{C}_T = \mathcal{C}_T^{(0)} + {\rm span} \{\partial_3,\,\partial_4\} $ where $\mathcal{C}_T^{(0)} $ is characterized by $\mathbf{Z}x^B =0$,  we have that\\
\centerline{$\mathcal{C}_T^{(0)} $ is a Lie algebra whose dimension is at most 3 and $f^B$ are arbitrary functions.}

\item[Case 2.H.2] If $K_{A|\alpha\beta} \neq 0$, $A=3, 4$, then
\begin{itemize}
\item either the system (\ref{MC3.H10c}) is Cramer's and $\mathcal{C}_T$ is a Lie algebra whose dimension is at most three 
\item or else, if the system (\ref{MC3.H10d}) is Cramer's, $\mathcal{C}_T$ is a Lie algebra whose dimension is at most five.
\end{itemize}
We have left unsolved the case when neither (\ref{MC3.H10c}) nor (\ref{MC3.H10d}) are Cramer's systems. This is a highly nongeneric instance that would require further study.  

\item[Case 2.H.1] $K_{4|\alpha\beta} = 0$ but $K_{3|\alpha\beta} \neq 0$, then we can expand: 
$\mathcal{C}_T = \mathcal{C}_T^{(0)} + {\rm span} \{\partial_4\}$ where $\mathcal{C}_T^{(0)} $ is characterized by $\mathbf{X}x^4 =0$.
Then either the system (\ref{MC3.H10e}) is Cramer's  and $\mathcal{C}_T^{(0)} $ is a Lie algebra whose dimension is at most three or else, the analogous of equation (\ref{MC3.H10d}) is a Cramer's system and $\mathcal{C}_T^{(0)} $ is a Lie algebra whose dimension is at most four. In both instances $f^4$ is an arbitrary function.
We have neither considered the nongeneric case in which neither (\ref{MC3.H10c}) nor (\ref{MC3.H10d}) are Cramer's systems.
\end{description}
Notice that, as cases {\bf 2.H.0} and {\bf 2.H.1} involve arbitrary functions, the corresponding classes of collineation fields are infinite dimensional and are not Lie algebras.
\end{description}

\section*{Acknowledgment}
The author is indebted to J Carot for calling his attention on the subject and for stimulating comments and discussions. Also the author is grateful for the warm hospitality of the Edinburgh Mathematical Physics Group (Edinburgh University), where a significant part of the present work was developed during a sabbatical stay. This work is supported by Ministerio de Educacion y Ciencia through grant no FIS2007-63034 and by Generalitat de Catalunya, 2009SGR-417 (DURSI).

%%%%%%%%%%%%%%%%%%%%%%%%%%%%%%%%%%%%%%%%%%%%%%%%%%%%%%%%%%%
\section*{Appendix: The derivation of equation (\ref{MC3.14}) }
To derive equation (\ref{MC3.14}) we apply $\nabla_\mu$ and $\partial_4$ respectively to equations (\ref{MC3.10}) and (\ref{MC3.7}) and then substract. Other facts that must be taken into account are that
$$ \partial_4 \nabla_\mu\Omega_\kappa^{\;\lambda} - \nabla_\mu \partial_4\Omega_\kappa^{\;\lambda} = \dot\Gamma^\lambda_{\mu\alpha} \Omega_\kappa^{\;\alpha} - \dot\Gamma^\alpha_{\mu\kappa} \Omega_\alpha^{\;\lambda} \,,$$
with 
$$\dot\Gamma^\alpha_{\mu\kappa}:= \partial_4\Gamma^\alpha_{\mu\kappa} = \nabla_\mu K^\alpha_{\;\kappa} + \nabla_\kappa K^\alpha_{\;\mu} - \nabla^\alpha K_{\mu\kappa}  \,,$$
and that:
\begin{eqnarray*}
\partial_4\nabla^\lambda\left(f K_{\mu\kappa} \right) & = & \nabla^\lambda\partial_4 \left(f K_{\mu\kappa} \right) - 2 K^{\alpha\lambda} \nabla_\alpha \left(f K_{\mu\kappa} \right) - f T^{\alpha \lambda} \left(\dot\Gamma^\nu_{\alpha\mu} K_{\nu\kappa} + \dot\Gamma^\nu_{\alpha\kappa} K_{\mu\nu} \right) \\
\partial_4\nabla_\kappa\left(f K^\lambda_{\;\mu} \right) & = & \nabla_\kappa\partial_4\left(f K^\lambda_{\;\mu} \right) + f \left( \dot\Gamma^\lambda_{\kappa\alpha} K^\alpha_{\;\mu} - \dot\Gamma^\alpha_{\mu\kappa} K^\lambda_{\;\alpha} \right)
\end{eqnarray*}

In case that $\dot K^\alpha_{\;\mu} = b K^\alpha_{\;\mu}$, it follows that 
$$ \partial_4\left(f K^\alpha_{\;\mu} \right) = (\dot f + b f ) K^\alpha_{\;\mu} 
\qquad {\rm and} \qquad \partial_4\left(f K_{\mu\kappa} \right) = (\dot f + b f ) K_{\mu\kappa} + 2 f K_{\alpha\kappa}K^\alpha_{\;\mu}  $$

Using all that, after a little algebra we finally obtain 
$$ f_{|\alpha} \left(K^{\lambda\alpha} K_{\mu\kappa} - K_{\mu\beta} K^\beta_{\;\kappa} T^{\alpha\lambda} \right) = W^\lambda_{\mu\kappa} $$
with  
\begin{eqnarray} 
W^\lambda_{\mu\kappa} &:=& - \nabla_{(\mu} \left[(\dot f + b f ) K^\lambda_{\;\kappa)}\right] + \frac12\,\nabla^\lambda \left[ (\dot f + b f ) K_{\mu\kappa}\right] + \frac12\, Z^\alpha \left(\dot R_{\alpha\mu\kappa}^{\hspace*{1em}\;\lambda} - \nabla_\mu \dot\Gamma^\lambda_{\kappa\alpha}    \right) +\dot\Gamma^\lambda_{\alpha(\mu}\Omega^\alpha_{\;\kappa)} - \nonumber \\
 & & \frac12\,\dot\Gamma^\alpha_{\mu\kappa}\Omega^\lambda_{\;\alpha} - f \left( T^{\alpha\lambda} \dot\Gamma^\nu_{\alpha(\mu} K_{\nu\kappa)}  - \frac12\,\dot\Gamma^\nu_{\mu\kappa} K^\lambda_{\;\nu}  +  K^{\alpha\lambda} \nabla_\alpha K_{\mu\kappa} - \nabla^\lambda \left[K_{\alpha\kappa}K^\alpha_{\;\mu} \right] \right)  \label{MC3.20} 
\end{eqnarray}
which,  as  $\dot f + b f$ is a function of $\mathbf{Z}$, is a linear function of $Z^\alpha$, $\Omega_{\mu\nu}$ and $f$.

\end{document}